\documentclass[useAMS,usenatbib]{mn2e}

\usepackage{pdflscape}
\usepackage{epstopdf}
\usepackage{graphicx}
\usepackage{textcomp}
\usepackage{courier}
\usepackage{caption}
\usepackage{amssymb,amsmath}
\usepackage{url}
\usepackage{deluxetable}

\newcommand{\aj}{AJ}
\newcommand{\apj}{ApJ}
\newcommand{\apjl}{ApL}
\newcommand{\apjs}{ApJS}
\newcommand{\iaucirc}{IAU Circ.}
\newcommand{\mnras}{MNRAS}
\newcommand{\araa}{ARA\&A}
\newcommand{\nat}{Nature}
\newcommand{\pasp}{PASP}
\newcommand{\aap}{A\&A}
\makeatletter
\def\blfootnote{\xdef\@thefnmark{}\@footnotetext}
\makeatother

\title[Constraints on SN IIn Progenitor Outbursts from LOSS]{Constraints on Type IIn Supernova Progenitor Outbursts from the Lick Observatory Supernova Search}
\author[Bilinski et. al.]{Christopher Bilinski$^{1}$\thanks{E-mail:
cgbilinsk@gmail.com}, Nathan Smith$^{1}$, Weidong Li$^{2}$\thanks{Deceased 2011 December 12}, G. Grant Williams$^{3}$,
\newauthor WeiKang Zheng$^{2}$, and Alexei V. Filippenko$^{2}$\\
$^{1}$Steward Observatory, University of Arizona, 933 N. Cherry Avenue, Tucson AZ 85721, USA\\
$^{2}$Department of Astronomy, University of California, Berkeley, CA 94720-3411, USA\\
$^{3}$MMT Observatory, Tucson, AZ 85721-0065, USA}
\begin{document}

\date{Accepted 2015 March 12.  Received 2015 February 25; in original form 2014 October 13}

\pagerange{\pageref{firstpage}--\pageref{lastpage}} \pubyear{2015}

\maketitle

\label{firstpage}

\begin{abstract}
We searched through roughly 12 years of archival survey data acquired by the Katzman Automatic Imaging Telescope (KAIT) as part of the Lick Observatory Supernova Search (LOSS) in order to detect or place limits on possible progenitor outbursts of Type IIn supernovae (SNe~IIn).  The KAIT database contains multiple pre-SN images for 5 SNe~IIn (plus one ambiguous case of a SN IIn/imposter) within 50 Mpc.  No progenitor outbursts are found using the false discovery rate (FDR) statistical method in any of our targets.  Instead, we derive limiting magnitudes (LMs) at the locations of the SNe.  These limiting magnitudes (typically reaching $m_R \approx 19.5\,\mathrm{mag}$) are compared to outbursts of SN 2009ip and $\eta$ Car, plus additional simulated outbursts.  We find that the data for SN 1999el and SN 2003dv are of sufficient quality to rule out events $\sim40$ days before the main peak caused by initially faint SNe from blue supergiant (BSG) precursor stars, as in the cases of SN 2009ip and SN 2010mc.  These SNe~IIn may thus have arisen from red supergiant progenitors, or they may have had a more rapid onset of circumstellar matter interaction.  We also estimate the probability of detecting at least one outburst in our dataset to be $\gtrsim60\%$ for each type of the example outbursts, so the lack of any detections suggests that such outbursts are either typically less luminous (intrinsically or owing to dust) than $\sim -13\,\mathrm{mag}$, or not very common among SNe~IIn within a few years prior to explosion.
\end{abstract}

\begin{keywords}
supernovae: Type IIn --- luminous blue variables --- supernova progenitors
\end{keywords}

\section{Introduction}
\label{sec:Int}
Core-collapse supernovae (CCSNe) exhibit a wide diversity in their light curves and spectral properties, and it remains an enduring challenge to connect these properties to the evolution of their progenitors.  Prior to 2009, fourteen CCSN progenitor stars or progenitor outbursts (eight SNe~II-P, three IIb, one IIn, one Ibn, and one II-pec) had been detected in pre-explosion images \citep[and references therein]{2009ARA&A..47...63S}.  Over the last few years, at least six more (IIn: 1961V, 2009ip, 2010jl; IIb: 2011dh, 2013df; IIn-P: 2011ht) have been found \citep{2010AJ....139.1451S,2011MNRAS.415..773S,2011ApJ...732...63S,2011ApJ...737...76K,2011ApJ...732...32F,2011ApJ...739L..37M,2014AJ....147...37V,2013ApJ...779L...8F}. Detection of these progenitors is crucial for gaining a better understanding of the observed SN diversity.  Some theoretical models for the end stages of massive-star evolution suggest that stars with $\sim 30$--60\,M$_{\odot}$ will collapse to form a black hole without a SN explosion \citep[][but see \citealp{2013ApJ...762..126O}]{1999ApJ...522..413F,2003ApJ...591..288H}.  The detection of what seem to be massive luminous blue variable stars (LBVs) as the progenitors of the Type IIn SNe 1961V \citep[]{2011ApJ...732...63S,2011ApJ...737...76K}, 2005gl \citep[]{2007ApJ...656..372G,2009Natur.458..865G}, 2009ip \citep[]{2010AJ....139.1451S,2011ApJ...732...32F}, and 2010jl \citep[]{2011ApJ...732...63S}, however, challenge this notion.

Optical spectra of SNe~IIn exhibit prominent, relatively narrow hydrogen emission lines which arise from the interaction of the expanding SN debris with a recently ejected circumstellar shell that is typically moving at $\lesssim 1000$ km $\mathrm{s^{-1}}$ \citep{1990MNRAS.244..269S,1997ARA&A}.  Because these lines appear soon after the SN explosion, it is expected that the shell of circumstellar matter (CSM) was ejected by the star within a few years before the SN explosion itself \citep[]{2014ARA&A..52..487S}.  Such pre-SN eruptions in the years just before the SN suggest that instabilities arise during the late stages of nuclear burning \citep{2014ApJ...785...82S,2012MNRAS.423L..92Q} to power mass-loss episodes that might result in a detectable brightening of the progenitor system.  In fact, there have already been a few direct detections of pre-SN outbursts: SN 2006jc \citep[Ibn;][]{foley07,2007Natur.447..829P}, SN 2009ip \citep[IIn;][]{2010AJ....139.1451S,2013ApJ...767....1P,2013MNRAS.430.1801M}, SN 2011ht \citep[IIn-P;][]{2013ApJ...779L...8F}, and arguably SN 1961V \citep[IIn;][]{2011ApJ...732...63S,2011ApJ...737...76K}, thus strengthening the theoretical connection between LBVs and SNe~IIn.\footnote{Note that we distinguish between a detection of a progenitor and a detection of variability that indicates a pre-SN outburst.  It remains possible that some SNe with only one epoch of a detected progenitor may have in fact been detected during a relatively bright outburst phase.}  Even when no direct detection has been made, inferred wind and mass-loss parameters \citep[]{2007ApJ...666.1116S,2007ApJ...656..372G,2008ApJ...686..467S,2010ApJ...709..856S,2012ApJ...744...10K,2013A&A...555A..10T,2014ARA&A..52..487S}, strong infrared excesses at late times \citep[]{2002ApJ...575.1007G,2011ApJ...741....7F}, and a theoretical connection to quasi-periodic radio modulations \citep[]{2006A&A...460L...5K} suggest that LBV-like progenitors are likely for some SNe~IIn.  \citet{2014arXiv1401.5468O} estimate that $>50\%$ of SNe~IIn in their Palomar Transient Factory (PTF) data have pre-explosion outbursts brighter than $M=-14\,\mathrm{mag}$, though some of these pre-peak brightening events may actually be the SN explosion itself (this is discussed more in \S\ref{sec:RvB}).

Recent work suggests that SNe~IIn follow LBV-like eruptions, although it remains unclear if the driving instability of the pre-SN eruptions is the same as the instability of classical LBVs \citep[see review by][]{2014ARA&A..52..487S}.  In the context of stellar evolution, LBVs were suggested to be a transitional phase between O-type stars and Wolf-Rayet stars that skipped a red supergiant phase.  Giant eruptions, occurring when a massive star increases its bolometric luminosity \citep[]{1999PASP..111.1124H,1994PASP..106.1025H}, were included in the LBV phenomenon.  By spatially resolving the circumstellar shells resulting from two such LBV outbursts within our own Galaxy (P Cygni and $\eta$ Carinae), studies have been able to measure the amount of mass lost in observed LBV giant eruptions.  Estimates of P Cygni's circumstellar shell from the 1600 AD outburst total only about 0.1\,M$_{\odot}$ \citep[]{2006ApJ...638.1045S}, whereas estimates of $\eta$ Carinae's shell from its mid-19th century outburst total around 10--20\,M$_{\odot}$ \citep[]{2003AJ....125.1458S,2007ApJ...655..911S,2010MNRAS.401L..48G}.  While other Galactic LBVs have been observed with CSM shells surrounding them \citep{2006ApJ...645L..45S}, only these two events were actually observed during the eruptions that produced their CSM shells. 

A number of recent developments have challenged the traditional view of LBVs, including evidence of shock-powered events rather than super-Eddington winds \citep{2008Natur.455..201S,2013MNRAS.429.2366S}, light-echo spectra that do not match expectations for winds \citep{2012Natur.482..375R,2014ApJ...787L...8P}, and a widening range of initial masses that experience outbursts \citep{2008ATel.1550....1P,2008ApJ...681L...9P,2009ApJ...705.1364T}.  \citet[]{2014arXiv1406.7431S} show that LBVs tend to be more isolated than O-type stars and even Wolf-Rayet stars, challenging the idea that they are in transition between these two phases.  Instead, \citet[]{2014arXiv1406.7431S} propose that they may be the late evolutionary stage of mass gainers or mergers in binary systems.  Finally, as noted above, the discovery of a number of LBVs exploding as CCSNe without first experiencing a relatively long ($>0.5$\,My) Wolf-Rayet phase also challenges the traditional view of LBVs.

In this paper, we further explore the connection between LBVs and SNe~IIn by searching for additional progenitor outbursts preceding the SNe~IIn. The data were acquired starting in 1998 as part of the Lick Observatory Supernova Search (LOSS) with the 0.76\,m Katzman Automatic Imaging Telescope \citep[KAIT;][]{filippenko01}.  KAIT began automatic operation and discovered its first SN in 1997 \citep[]{1997IAUC.6627....1T}.  While the program was primarily focused on the determination of SN rates \citep{leaman2011,li2011b,li2011c} and SNe~Ia for use as cosmological probes \citep{2010ApJS..190..418G,2011MNRAS.416.2607G}, the KAIT database includes pre-SN images and SN photometry of many SN events that can be utilised for other purposes \citep{2011Natur.480..348L}.  SN 1997bs, the first object that KAIT discovered with a SN designation, was interpreted as not being a SN at all, and became a prototype for the class of ``SN imposters'' \citep{2000PASP..112.1532V}.  A recent study by \citet{2015arXiv150200001A}, however, raises some additional questions about whether or not this was a terminal explosion.  The discussion of SN IIn progenitors has intensified because of the recent discovery of erupting progenitor systems like SN 2009ip \citep{2010AJ....139.1451S,2011ApJ...732...32F,2013ApJ...767....1P}.  

Here we aim to constrain the frequency and luminosity of such outbursts.  Information about the KAIT/LOSS program and the targets selected for our study is presented in \S \ref{sec:Obs}.  An analysis of our statistical approach to searching for progenitor outbursts and then constraining the brightness of these events with limiting magnitudes is given in \S \ref{sec:Met}. Section \ref{sec:Dis} discusses the nature of the progenitors and our fractional coverage rates. We conclude in \S \ref{sec:Con} with implications for future searches.

\section{Observations and Data Analysis}
\label{sec:Obs}
\subsection{Data Acquisition, Reduction, and Processing}
Unfiltered photometry of over 850 SNe has been obtained since 1998 with KAIT (0.76\,m) at Lick Observatory.  Over time, the telescope has used three different CCDs, each with a response that when combined with the optical path efficiencies results in a sensitivity most similar to that of the $R$ band \citep{2003PASP..115..844L,2010ApJS..190..418G}.  For this reason, we do all of our photometric analysis and comparison in $R$ when possible.  With a scale of $0\farcs8$ pixel$^{-1}$ and an effective chip size of $500\times500$ pixels, KAIT's total field of view is $6\farcm7\times6\farcm7$ \citep{2003PASP..115..844L}.  Typical integration times for KAIT span the range of 16--40\,s. During much of the time since 1998, KAIT has monitored roughly 15,000 galaxies at redshifts $z<0.05$ in search of transients.  Over 1200 galaxies can be observed in a single long winter night (but fewer than 800 in a short summer night), resulting in a cadence of roughly 3--10 days for most targets (excluding the portion of the year that the target cannot be observed because it is in the daytime or bright twilight sky).  These transient candidates go through a process of being automatically compared to template images and flagged, verified by human checkers, and then reobserved the following night before a SN is announced \citep[]{filippenko01,2011MNRAS.416.2607G,leaman2011}.

Reduction and processing of the KAIT images was performed using techniques described by \citet[]{2011Natur.480..348L}.  Bias corrections and flatfielding are automatically performed at the telescope for each target.  The primary template image used for alignment and galaxy subtraction was obtained as part of the routine observation program on a photometric night with a longer exposure time than the typical images (usually 40\,s as opposed to 16--30\,s).  A deep template with the SN present was generated using all images up to one year after the SN explosion that were of high quality and still contained the SN event as verified visually after galaxy subtraction.  This deep SN template was used to provide precise astrometry of the SN for the later analysis using artificial star injection.

We obtained astrometric solutions for our primary template image by uploading our image to Astrometry.net \citep[]{2008AJ....135..414B}.  We then applied these astrometric solutions to the data images after they were aligned with the template image.  The IRAF tasks \texttt{phot}, \texttt{psf} (on one of the brightest stars away from the galaxy or edge of the image), and \texttt{allstar} in the DAOPHOT package provided aperture and point-spread-function (PSF) photometry for all of our images.  We calibrated the measured magnitudes to known magnitudes for any stars also present in the USNO-B database, which has rather large uncertainties of about 0.25 mag \citep[]{2003AJ....125..984M}.  We performed image subtraction using the \texttt{hotpants} program\footnote{\texttt{hotpants} is available at \url{http://www.astro.washington.} \url{edu/users/becker/v2.0/hotpants.html}.}, after which we could use the DAOPHOT package \texttt{addstar} task to inject artificial stars.  The \texttt{phot} task was used again to measure the photometry of the artificial stars and confirm that these artificial stars in fact matched the known magnitudes that we inserted.  We combined our images by using the \texttt{imcombine} task in IRAF in average mode when needed. The combined images were often cropped in order to obtain an intersection of our data that excluded regions which only contained data from a number of images less than the total in the stack.

\subsection{Type IIn Supernova Targets}
\label{sec:List}

From the large sample of KAIT-observed SNe, we first selected all SNe~IIn, candidate SNe~IIn, and SNe~II as classified in the IAU Central Bureau of Astronomical Telegrams (CBAT) list of SNe.  We then restricted this sample to targets within 50 Mpc because KAIT images having a limiting apparent magnitude of $m_R \approx 19.5$ result in an absolute magnitude of $M_R \approx -14$ at this distance (neglecting reddening).  This is the peak luminosity of $\eta$ Carinae's 19th-century eruption to which we compare the progenitor outburst limiting magnitudes, and it is a typical value for SN imposters \cite[although they exhibit a wide range;][]{2011MNRAS.415..773S}.  The classification of the objects in the IAU CBAT list of SNe is usually based on the first announcement published, but these are often revised upon further study.  Accordingly, we checked the literature for each of our targets to verify that they were indeed SNe~IIn.  Overall, this selection process provided us with five unambiguous SNe~IIn and one ambiguous SN~IIn/SN imposter (SN 2006am).    

\begin{figure*}
\centering
\includegraphics[width=0.4\textwidth,clip=true,trim=0cm 0cm 0cm 0cm]{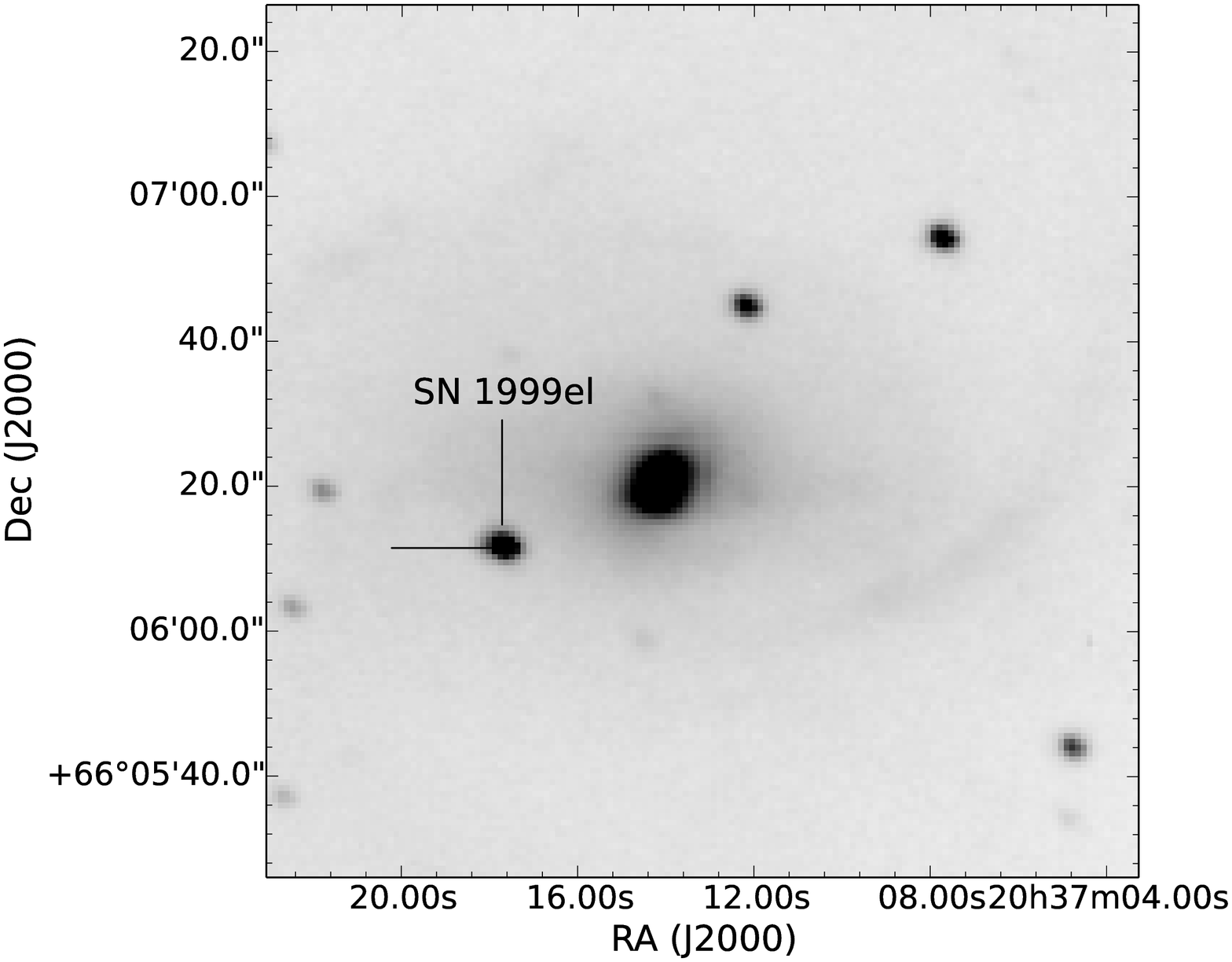}
\includegraphics[width=0.40\textwidth,clip=true,trim=0cm 0cm 0cm 0cm]{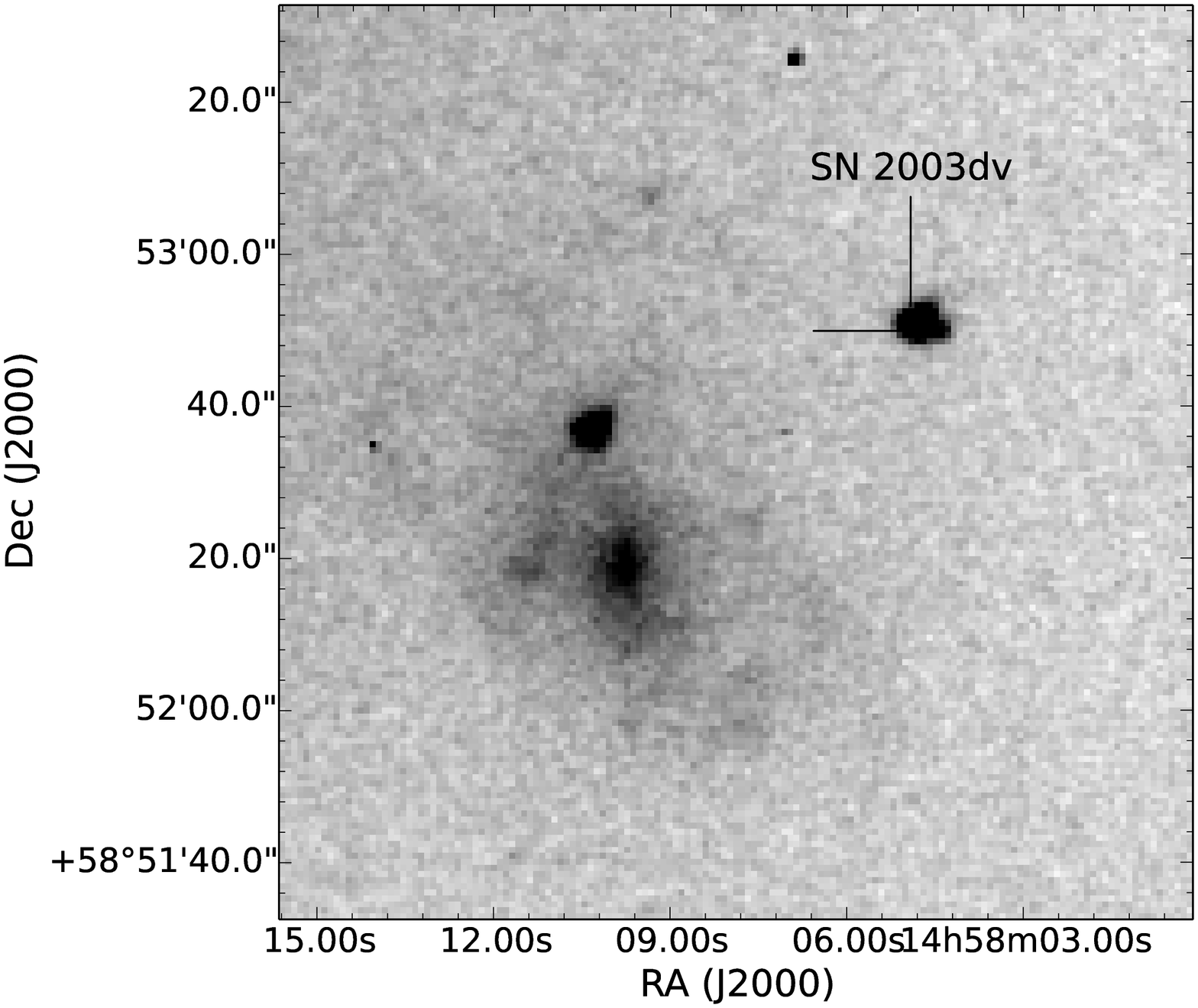} \\
\includegraphics[width=0.40\textwidth,clip=true,trim=0cm 0cm 0cm 0cm]{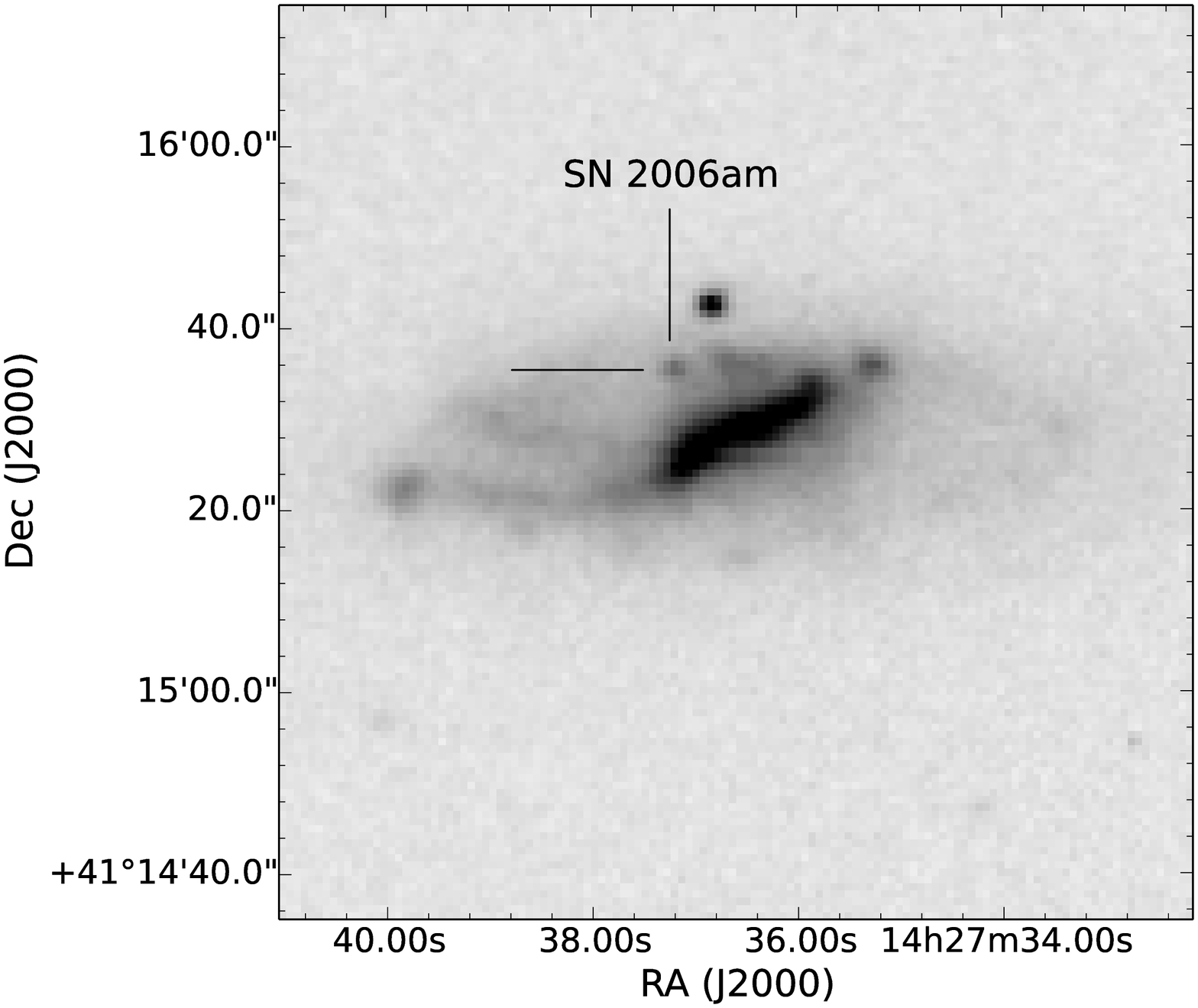} 
\includegraphics[width=0.4\textwidth,clip=true,trim=0cm 0cm 0cm 0cm]{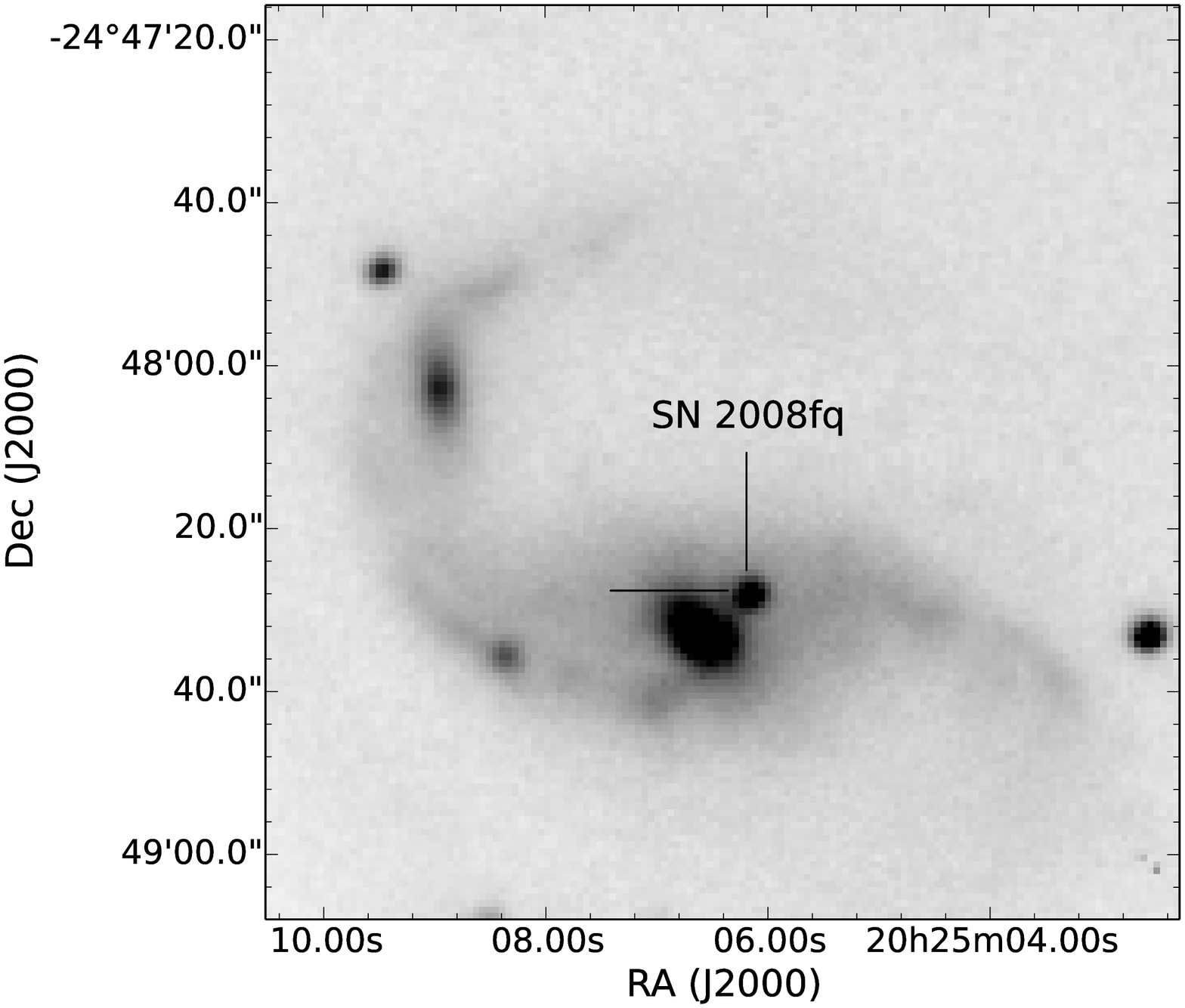} \\
\includegraphics[width=0.4\textwidth,clip=true,trim=0cm 0cm 0cm 0cm]{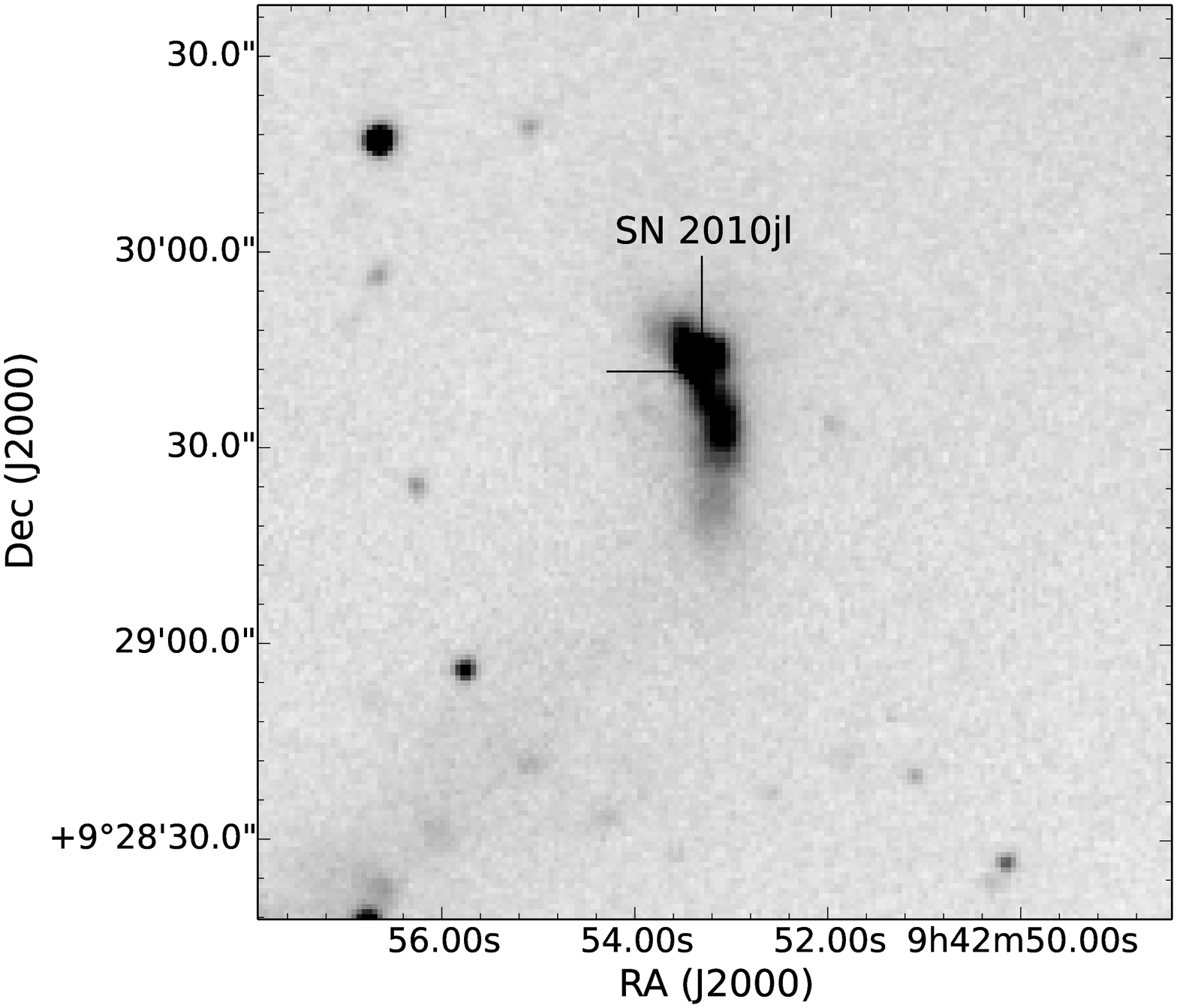}
\includegraphics[width=0.4\textwidth,clip=true,trim=0cm 0cm 0cm 0cm]{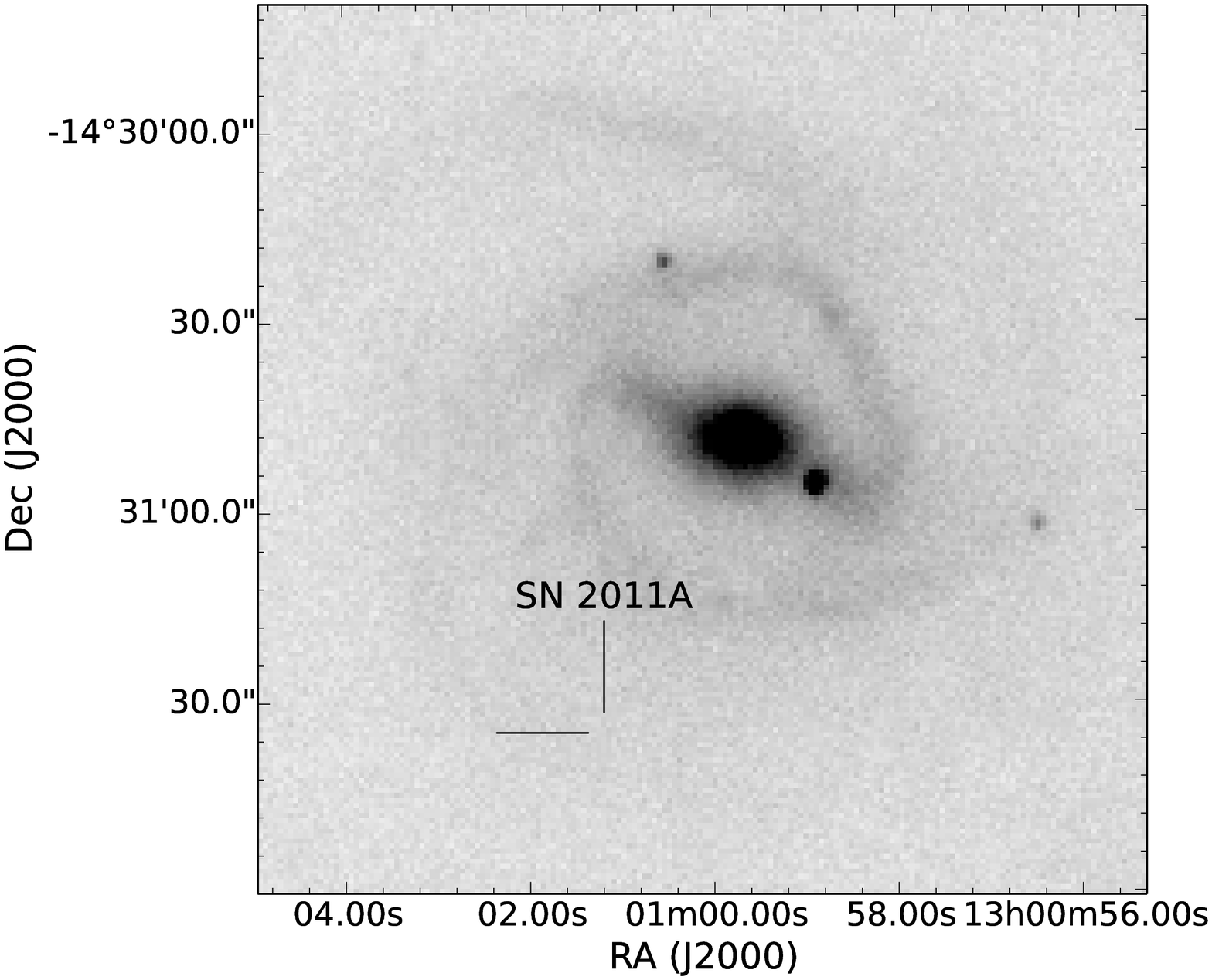}
\caption{Stacked images for each of the SNe~IIn showing the SN location.  Each image contains the actual SN except that of SN 2011A because no images were taken with the SN present.  For SN 2011A, the tick marks indicate where the SN would have been had we obtained post-SN images of this region.}
\label{fig:SNimages}
\end{figure*}

Figure \ref{fig:SNimages} shows stacked images of each SN field with the SN location labeled.  Table \ref{tab:SNTargets} enumerates many of the important properties of each of the targets. In all but one case (SN 1999el), use of Hubble's law was the only method available for distance determination for our targets.  We obtain these redshift-based distances from the NASA/IPAC Extragalactic Database\footnote{The NASA/IPAC Extragalactic Database (NED) is operated by the Jet Propulsion Laboratory, California Institute of Technology, under contract with the National Aeronautics and Space Administration (\url{http://ned.ipac.caltech.edu}).}, which assumes $H_0=73\,\mathrm{km\,s^{-1}\,Mpc^{-1}}$ \citep{2005ApJ...627..579R} and takes into account influences from the Virgo cluster, the Great Attractor, and the Shapley supercluster.  Extinction along the line of sight to each SN's host galaxy due to the Milky Way was applied \citep{2011ApJ...737..103S}.  Extinction produced in the host galaxies themselves is not known for any of our targets.  Since these SNe~IIn show signs of prior mass loss resulting in CSM, it is possible that the progenitors were enshrouded in dusty CSM that was destroyed by the SN.  Our limiting magnitudes do not account for this possible additional extinction.

\begin{table*}
\begin{minipage}{200mm}
\caption{List of Type IIn Supernova Targets and their Properties}
\label{tab:SNTargets}
\begin{tabular}{cccccccc}
  \hline
	SN & Host Galaxy & UT Peak Date & RA (J2000)\tablenotemark{a} & Dec (J2000) & Dist. (Mpc)\tablenotemark{b} & Peak App. Mag. (Band) & Peak Abs. Mag. \\
	\hline
	1999el & NGC 6951 &	Nov-08-1999	&	20$^h$37$^m$17.72$^s$	&	$+66^\circ 06' 11.5''$	&	26.3\tablenotemark{c}	&	14.5 (R) & -18.4 \\
	2003dv & UGC 9638 &	May-04-2003	&	14 58 04.92	&	+58 52 49.9	&	38.0 & 15.4 (uf/R) & -17.5 \\
	2006am & NGC 5630 &	Feb-24-2006	&	14 27 37.24	&	+41 15 35.4	&	44.2 &	17.0 (uf/R) & -16.2	\\
	2008fq & NGC 6907 & Sep-22-2008 & 20 25 06.19 & -24 48 27.6 & 47.4 & 15.4 (uf/R) & -18.1 \\
	2010jl & UGC 5189A &	Nov-05-2010	&	09 42 53.33	&	+09 29 41.8	&	48.9\tablenotemark{d} &	12.9 (uf) & -20.6	\\
	2011A & NGC 4902 &	Jan-2-2011	&	13 01 01.19	&	-14 31 34.8	&	38.2 &	16.9 (uf) & -16.1	\\
	\hline
	\end{tabular}
\tablenotetext{a}{\raggedright Peak dates are based off of published data if a well-sampled light curve exists.  In the case that the published data are limited, we instead determine the peak date from the KAIT data.}
\tablenotetext{b}{\raggedright Unless specifically cited, distances were obtained from the NED database, which assumes $H_0=73\,\mathrm{km\,s^{-1}\,Mpc^{-1}}$ and takes into account influence from the Virgo cluster, the Great Attractor, and the Shapley supercluster.  Milky Way extinction values along the line of sight to the host galaxies were taken into account when determining absolute magnitudes \citep[]{2011ApJ...737..103S}.}
\tablenotetext{c}{\raggedright\citet[]{2002ApJ...573..144D}.}
\tablenotetext{d}{\raggedright\citet[]{2011ApJ...732...63S}.}
\end{minipage}
\end{table*}

\subsubsection{SN 1999el}
SN 1999el was discovered near NGC 6951 as a part of the Beijing Astronomical Observatory Supernova Survey (BAOSS; \citealp{1999IAUC.7288....1C}) and peaked at  $m_R = 14.5\,\mathrm{mag}$ ($M_{R}=-18.4\,\mathrm{mag}$; \citealp{2002ApJ...573..144D}) on 8 November 1999 (UT dates are used throughout this paper).  It was determined to be a SN~IIn based on BAO spectra \citep[]{1999IAUC.7288....1C}.  Follow-up observations in the near-infrared also confirm its classification as a SN~IIn \citep[]{2001A&AT...20..385D, 2002ApJ...573..144D, 1999IAUC.7303....1D}.  The ultraviolet/optical luminosity resulting from interaction between the SN shock and pre-existing circumstellar dust caused a blueward shift of the $J$, $H$, and $K$ light curves within 5--80 days after discovery \citep{2001A&AT...20..385D}.  Light echoes from hot CSM dust are also thought to play a role in the near-infrared light curves observed over the course of $\sim416$ days following maximum light \citep{2002ApJ...573..144D}.  These signs of interaction between the SN explosion and CSM are likely the result of mass-loss episodes prior to the final explosion.  The distance to SN 1999el (26.3\,Mpc) was determined by using the Type Ia SN 2000E in NGC 6951 \citep{2002ApJ...573..144D}.  A Galactic extinction value of $A_R=0.808\,\mathrm{mag}$ was adopted from \citet{2011ApJ...737..103S}.

\subsubsection{SN 2003dv}
SN 2003dv was discovered near UGC 9638 as part of the Lick Observatory and Tenagra Observatory Supernova Search \citep[LOTOSS;][]{2003IAUC.8124....1K} and peaked at an unfiltered apparent magnitude of 15.4 ($M_{R}=-17.5\,\mathrm{mag}$) on 4 May 2003\footnote{Light curve taken from \url{http://www.rochesterastronomy.} \url{org/sn2003/sn2003dv.html}.}.  Spectra taken about a week before maximum brightness suggest that it was a SN~IIn based on its narrow ($350\,\mathrm{km\,s^{-1}}$) H$\alpha$ profile, which is slightly redward of the peak of the broad component ($10,000\,\mathrm{km\,s^{-1}}$) of this Balmer emission \citep[]{2003IAUC.8124....1K}.  The distance to UGC 9638 obtained from NED as described in \S\ref{sec:List} is 38.0 Mpc.  A Galactic extinction of $A_R=0.030\,\mathrm{mag}$ was adopted from \citet{2011ApJ...737..103S}.

\subsubsection{SN 2006am}
SN 2006am was discovered near NGC 5630 as a part of LOSS with an unfiltered apparent magnitude of 18.5 ($M_{R}=-14.8\,\mathrm{mag}$) on 22 February 2006 \citep[]{2006CBET..412....1L}.  This absolute magnitude indicates a rather low luminosity for a CCSN, suggesting that SN 2006am may be either a SN imposter or a significantly extinguished SN.  However, SN 2006am exhibits properties of a SN~IIn with a blue continuum and Balmer emission lines containing both a narrow component (300--400\,$\mathrm{km\,s^{-1}}$) and a weaker broad component (2000\,$\mathrm{km\,s^{-1}}$; \citealp[]{2006IAUC.8680....1L}).  Given its faint magnitude along with its blue continuum and narrow spectral lines, SN 2006am may well be a SN imposter --- but it may also be a SN~IIn, so we include it in our sample.  The distance to NGC 5630 obtained from NED as described in \S\ref{sec:List} is 44.2 Mpc.  A Galactic extinction of $A_R=0.025\,\mathrm{mag}$ was adopted from \citet{2011ApJ...737..103S}.

\subsubsection{SN 2008fq}
SN 2008fq was discovered near NGC 6907 as a part of LOSS with an unfiltered apparent magnitude of 15.4 ($M_{R}=-18.1\,\mathrm{mag}$) on 15 September 2008 \citep{2008CBET.1507....1T}.  Spectra showed narrow and blueshifted ($670\,\mathrm{km\,s^{-1}}$) P-Cygni-like H$\alpha$ and H$\beta$ components superposed on a blueshifted ($7500\,\mathrm{km\,s^{-1}}$) diffuse absorption feature \citep{2008CBET.1510....1Q}.  Likewise, the Na I line shows a similarly complicated profile.  \citet{2008CBET.1510....1Q} suggest the presence of multiple shells in the SN ejecta arising from ejecta-CSM interaction.  Although originally classified as a Type II SN, the narrow H$\alpha$ features overlaying a broader component probably indicate that this object was a SN~IIn.  The distance to NGC 6907 obtained from NED as described in \S\ref{sec:List} is 47.4 Mpc.  A Galactic extinction of $A_R=0.137\,\mathrm{mag}$ was adopted from \citet{2011ApJ...737..103S}.

\subsubsection{SN 2010jl}

SN 2010jl was discovered near UGC 5189A by \citet{2010CBET.2532....1N} and peaked at an unfiltered apparent magnitude of 12.9 ($M = -20.6\,\mathrm{mag}$) on 5 November 2010 \citep{2011ApJ...732...63S}.  This absolute magnitude places SN 2010jl in the class of superluminous supernovae (SLSNe) along with SN 2003ma \citep{2011ApJ...729...88R}, SN 2006tf \citep{2008ApJ...686..467S}, and SN 2006gy \citep{2007ApJ...659L..13O}. SN 2010jl's host galaxy, UGC 5189A, is reported to be moderately metal poor with $Z \approx 0.2$--0.5\,Z$_{\odot}$ \citep[]{2011ApJ...730...34S}.  Spectra taken two weeks after discovery exhibit a narrow component of $\sim120\,\mathrm{km\,s^{-1}}$ \citep{2011ApJ...732...63S}. 

Archival {\it Hubble Space Telescope (HST)} data reveal a luminous blue point source at the location of SN 2010jl approximately 10\,yr before its explosion \citep[]{2011ApJ...732...63S}.  Since the SN has not yet faded, this progenitor candidate could be a massive young star cluster, a quiescent luminous blue star, or a star undergoing an LBV-like eruption \citep[]{2011ApJ...732...63S}.  \citet{2014ApJ...781...42O} combined visible and X-ray data to estimate the presence of $>10\,{\rm M}_{\odot}$ of CSM surrounding the progenitor of SN 2010jl.  Because of its extreme brightness and relative proximity, SN 2010jl has been the focus of numerous further studies containing additional optical and X-ray observations \citep[]{2011ASInC...3..124R,2011ApJ...730...34S,2011A&A...527L...6P,2011AJ....142...45A,2012AJ....143...17S,2012ApJ...750L...2C,2012AJ....144..131Z,2013HEAD...1340004Z,2013arXiv1312.6617F}.  The distance to UGC 5189A obtained from NED as described in \S\ref{sec:List} is 48.9 Mpc, consistent with the value adopted by \citet{2011ApJ...732...63S}.  A Galactic extinction of $A_R=0.059\,\mathrm{mag}$ was adopted from \citet{2011ApJ...737..103S}.

\subsubsection{SN 2011A}
SN 2011A was discovered near NGC 4902 as a part of the Chilean Automatic Supernova search (CHASE) project at an unfiltered apparent magnitude of 16.9 ($M = -16.1\,\mathrm{mag}$) on 2 January 2011 \citep[]{2011CBET.2623....1P}.  Spectra were interpreted to suggest that SN 2011A was a SN~IIn similar to SN 2005cl \citep[]{2011CBET.2623....1P}.   The distance to NGC 4902 obtained from NED as described in \S\ref{sec:List} is 38.2 Mpc.  A Galactic extinction of $A_R=0.109\,\mathrm{mag}$ was adopted from \citet{2011ApJ...737..103S}.

\section{Analysis}
\label{sec:Met}
Our primary objective was to detect or place upper limits on the possible progenitor outbursts at the locations of the SNe in our sample, within the time frame monitored by KAIT prior to the SNe.  No progenitor outbursts were detected to a statistically significant level using the false discovery rate (FDR) statistical method (discussed in \S \ref{ssec:FDR})\footnote{We also searched for outbursts in SN~IIn imposter events such as SN 2001ac and SN 2006bv, but found no statistically significant detections in this sample either.}.  Therefore, we also used a number of different techniques to place limiting magnitude constraints for these SN progenitor outbursts as described below.

\subsection{False Discovery Rate Statistical Method}
\label{ssec:FDR}
The FDR statistical method \citep{FDRcite} was used to study the possibility of a source being detected in the archival KAIT images for each of the SN targets.  It controls for the fact that with a very large dataset, Gaussian fluctuations would result in some significant false detections if a naive hypothesis testing method was used even when no emission is truly present \citep[]{2011Natur.480..348L}.  In order to apply this method, we first compute a $p$-value for the flux level in each image at the location of the SN.  We then place the $p$-values in ascending order and attempt to find the largest $k$ such that $P_k\le\frac{k}{m}\alpha$, where $k$ is the index  of the $p$-value, $m$ is the total number of images considered, and $\alpha$ is the significance requirement. We take the null hypothesis to be that no emission was present at the locations of the SNe in the archival images.  All images with $k=i,...,k_{\rm largest}$ are declared as rejections of the null hypothesis, which means that they are considered detections of a source.  We chose a relatively low significance level of $\alpha=0.05$, which is roughly the $2\sigma$ detection threshold, in order to see if there were any weak detections we could explore further.  

With these techniques, we found no statistically significant sources of emission in any of our targets.  To verify that the statistical approach was working correctly, we also ran the FDR test on SN 2010jl images within two months after the SN explosion and were able to obtain significant detections in each image.  However, since we are not able to directly detect any progenitor outbursts, we focus instead on setting limiting magnitudes at the SN position for each of the targets and compare them to the well-known light curves of SN 2009ip and $\eta$ Carinae.  $\eta$ Carinae serves as an example of a nonterminal eruption over extended periods of time, whereas SN 2009ip serves as an example of a terminal event which had brief outbursts within the years prior to explosion.  Although the nature of SN 2009ip's 2012a event was initially debated \citep{2013ApJ...763L..27P, 2013ApJ...764L...6S, 2013ApJ...767....1P,2013MNRAS.430.1801M,2013MNRAS.433.1312F,2014ApJ...780...21M}, recent evidence in favor of the 2012a event being a terminal explosion has not been disputed since \citep[Smith, Mauerhan, \& Prieto 2014,][]{2014ApJ...787..163G,2014MNRAS.442.1166M}.  While both $\eta$ Carinae and SN 2009ip have extensive light curves that we can reference, we do not claim that all SNe~IIn have outbursts similar to those of these two events.  As such, we also compare the limiting magnitudes we set to an array of simulated outbursts that cover a wide range of parameter space in varying magnitudes, durations of outbursts, and number of outbursts in a given event.  

\subsection{Limiting Magnitudes on Individual Images}

\subsubsection{Artificial Star Injection}
We take two approaches to setting limiting magnitudes for our images.  Our first approach involves the use of artificial star injection.  After subtraction of a template image not containing the SN, we implant increasingly fainter stars in steps of 0.1\,mag at the location of the SN, starting from $m_R \approx 15\,\mathrm{mag}$ until we no longer detect a source present at the location of the SN to the 3$\sigma$ level ($m_{\rm err}<0.362\,\mathrm{mag}$) of our known inserted star.  

While this method should be the most realistic at setting a relevant limiting magnitude for our images, there are a few drawbacks.  In the case where the SN is in the brightest regions of its host galaxy, as is the case for SN 2008fq and SN 2010jl, subtraction errors become more significant, resulting in a lower threshold of sensitivity and a large increase in the standard deviation of our limiting magnitudes.  This technique is also susceptible to error from cosmic rays in the signal flux aperture.  Cosmic rays in the signal flux aperture cause systematically fainter limiting magnitudes to be set, though they are relatively infrequent in the data.  Cosmic rays present in the sky annulus used for noise considerations are controlled with a standard rejection procedure.  Lastly, we increase the magnitude of the artificial stars in increments of 0.1, causing some artificial scatter in our results owing to this discretisation.  Results from this approach are shown in Figures 2--7 as red triangles and given in Tables 2--7 as ``Artificial Star LM, Individual.''

\subsubsection{Sky Background Noise Calculations at the SN Location}
The second way in which we determine limiting magnitudes for KAIT images is by measuring the noise level in a bias-, dark-, and template-subtracted image in an annulus around the SN location and calculating the flux that would be required to produce a magnitude error of 0.362.  This allows us to solve for the limiting flux required to reach this signal-to-noise ratio (S/N) given the sky parameters {\it near} the SN, without actually using the flux at the SN location itself.  We convert this limiting flux to a limiting magnitude and compare the result to the artificial star injection method.  

Although this method is subject to galaxy-subtraction errors in the sky annulus, it is not susceptible to cosmic rays nor galaxy-subtraction artifacts in the signal flux aperture because it does not use the flux in this region in determining the limiting magnitude.  Consequently, it tends to produce less scatter.  Even though this approach does not use the flux information exactly at the location of the SN, it does use the same annulus for background noise information as the artificial star approach.  Results from this approach are given in Tables 2--7 as ``Background LM, Individual.''

\subsection{Limiting Magnitudes on Monthly Stacked Images}
Since we expect the signal from a progenitor outburst to be quite faint (i.e., $M_R \approx -14\,\mathrm{mag}$) , we also took the approach of averaging all of the images within the same month in order to obtain deeper stacked images.  These images were first registered to a template image in preparation for stacking.  After combination, we then ran the same artificial star injections and noise calculations at the location of the SNe.  The stacked images often result in deeper limiting magnitudes being set, but their temporal coverage restricts their utility in searching for outbursts on the $\sim10$ day timescale.  The primary utility of these stacked images is in providing coverage of the faintest ($M_R=-13\,\mathrm{mag}$) outbursts that we simulate.  Results from this approach are shown in Figures 2--7 as blue squares and given in Tables 2--7 as ``Artificial Star LM, Stacked'' and ``Background LM, Stacked.''

\subsection{Results}
\label{sec:Res}
As mentioned in \S \ref{ssec:FDR}, the FDR statistical approach resulted in no significant progenitor outburst detections.  Therefore, we focus in this section on limiting magnitudes and opt not to calculate precursor rates as \citet{2014arXiv1401.5468O} do because our data only allow us to place constraints on the properties of the possible outbursts.  Plots showing the limiting magnitudes for each of the targets are included in Figures 2--7. The absolute magnitude light curves of SN 2009ip \citep{2010AJ....139.1451S,2013ApJ...767....1P,2013MNRAS.430.1801M} and $\eta$ Carinae \citep{2011MNRAS.415.2009S} are included for comparison in these figures. The light curves shown are shifted in the time axis so that their peak observed magnitudes occur at $t=0$.  All of the limiting magnitude values for each of the targets are included in Tables 2--7.

\subsubsection{SN 1999el}
\begin{figure*}
\includegraphics[width=1\textwidth,clip=true,trim=0cm 0cm 0cm 0cm]{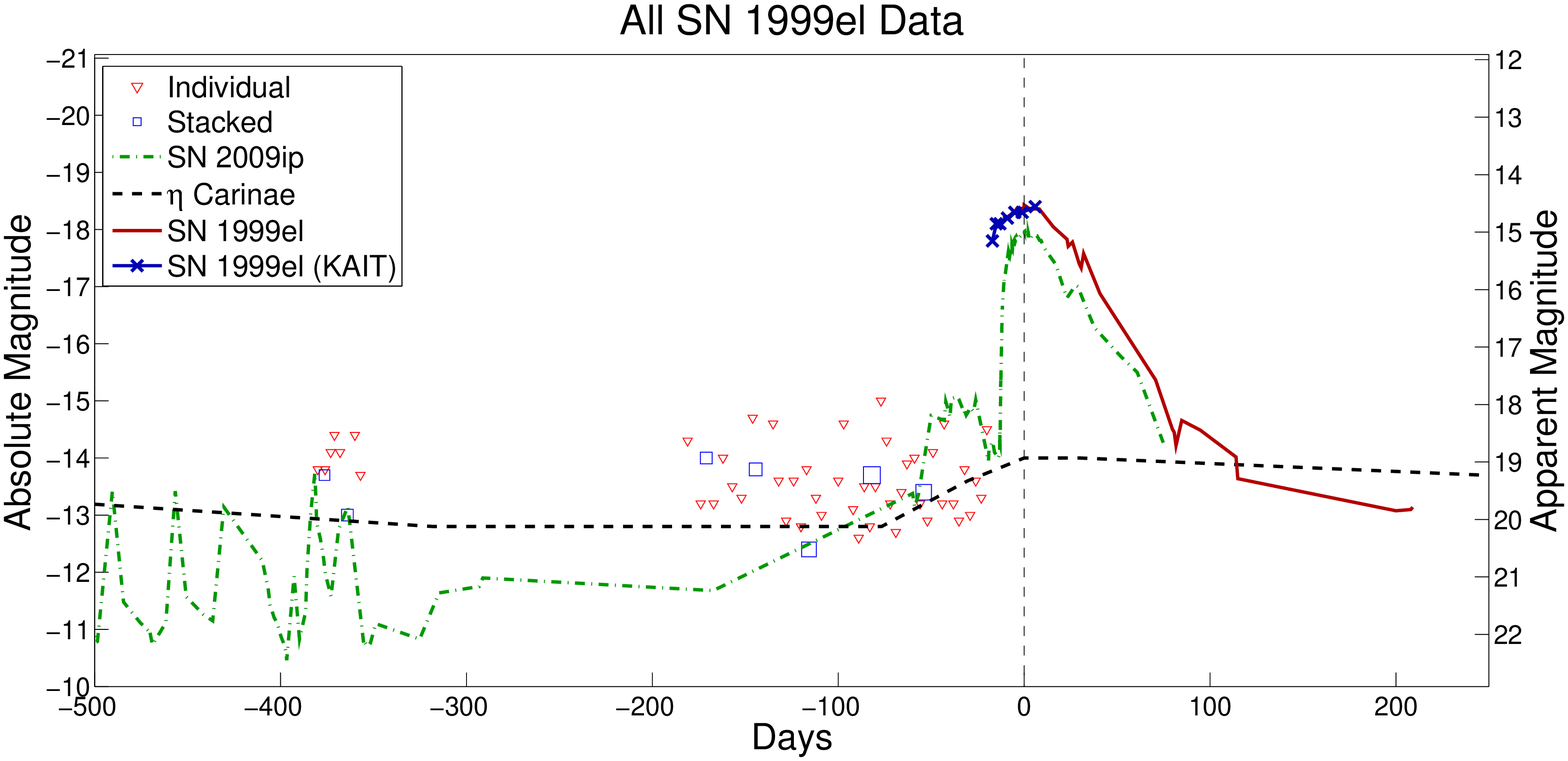} \\
\caption{Archival KAIT limiting magnitudes for SN 1999el.  Red triangular data signify limits set by data from single images, whereas blue-square data signify limits set by monthly stacks. (If only one image is available in a given month, the stack may consist of a single image resulting in overlap between the individual and combined limit.)  The size of the combined data squares indicates how many images went into producing the combined image (the largest stack consists of 16 images and belongs to the SN 2008fq dataset).  SN 2009ip's light curve is overplotted as a dash-dotted green line.  $\eta$ Carinae's light curve is overplotted as a dashed black line.  The vertical dashed black line marks the date of the peak observed magnitude.  Near-peak light-curve values for SN 1999el were obtained from \citet{2002ApJ...573..144D}.  We determine the date of the peak observed magnitude from this well-sampled light curve.}
\label{fig:LM1999el}
\end{figure*}

Results for SN 1999el are shown in Figure \ref{fig:LM1999el} and Table \ref{tab:LM1999el}.  While SN 1999el's dataset is relatively limited in duration because the SN exploded less than two years after KAIT began operation, it does contain many limits just before peak brightness and is the closest SN~IIn that we consider at 26.3 Mpc.  Consequently, the majority of the limiting magnitudes set for the SN 1999el data lie in the absolute magnitude range from $-13$ to $-14$ mag.  SN 1999el's light curve \citep{2002ApJ...573..144D} matches that of SN 2009ip (from the 2012b peak onward) very well both in magnitude and in shape.  SN 1999el was not observed as early in its rise to peak, but its post-peak light curve shows evolution similar to that of SN 2009ip.  As discussed below, we can rule out an initially faint SN if it were similar to SN 2009ip's 2012a event.

\begin{table*}
\begin{minipage}{200mm}
\caption{Limiting Magnitudes (LM) for SN 1999el\tablenotemark{a}}
\label{tab:LM1999el}
\hspace*{-30pt}
\begin{tabular}{ccccccc}
  \hline
	Year & Month & Day & Artificial Star LM, Individual & Artificial Star LM, Stacked & Background LM, Individual & Background LM, Stacked \\
	\hline
1998 & 10 & 24 & 19.2 & 19.2 & 19.4 & 19.8 \\
1998 & 10 & 28 & 19.1 & 19.2 & 19.6 & 19.8 \\
1998 & 10 & 31 & 18.8 & 19.2 & 19.7 & 19.8 \\
1998 & 11 & 2  & 18.5 & 19.9 & 19.6 & 19.8 \\
1998 & 11 & 5  & 18.8 & 19.9 & 18.8 & 19.8 \\
\hline
	\end{tabular}
	\hspace*{30pt}
\tablenotetext{a}{\raggedright The entirety of this table is available electronically.  A portion is displayed here for reference.}
\end{minipage}
\end{table*}

\subsubsection{SN 2003dv}
\begin{figure*}
\centering
\includegraphics[width=1\textwidth,clip=true,trim=0cm 0cm 0cm 0cm]{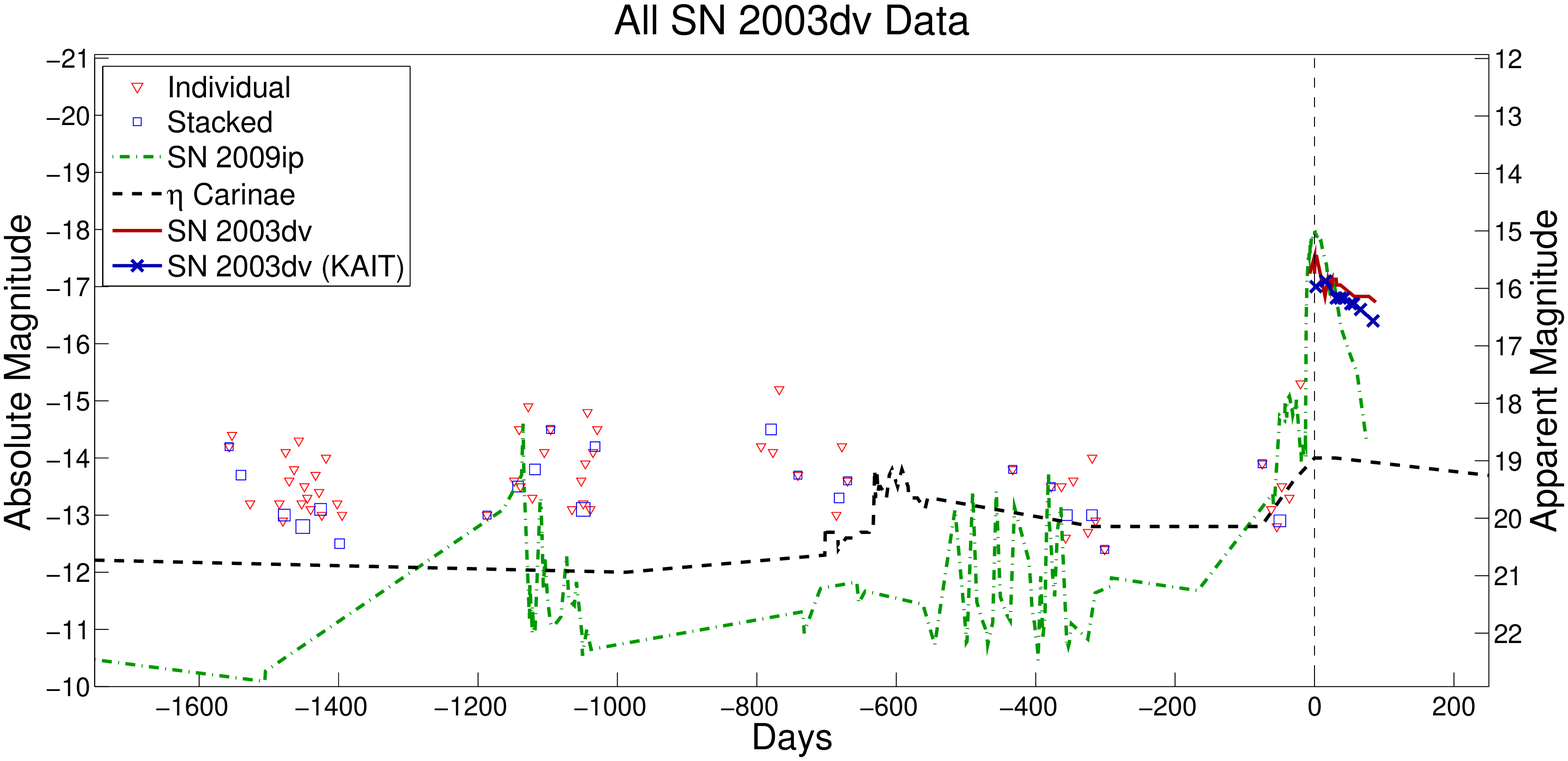} \\
\includegraphics[width=1\textwidth,clip=true,trim=0cm 0cm 0cm 0cm]{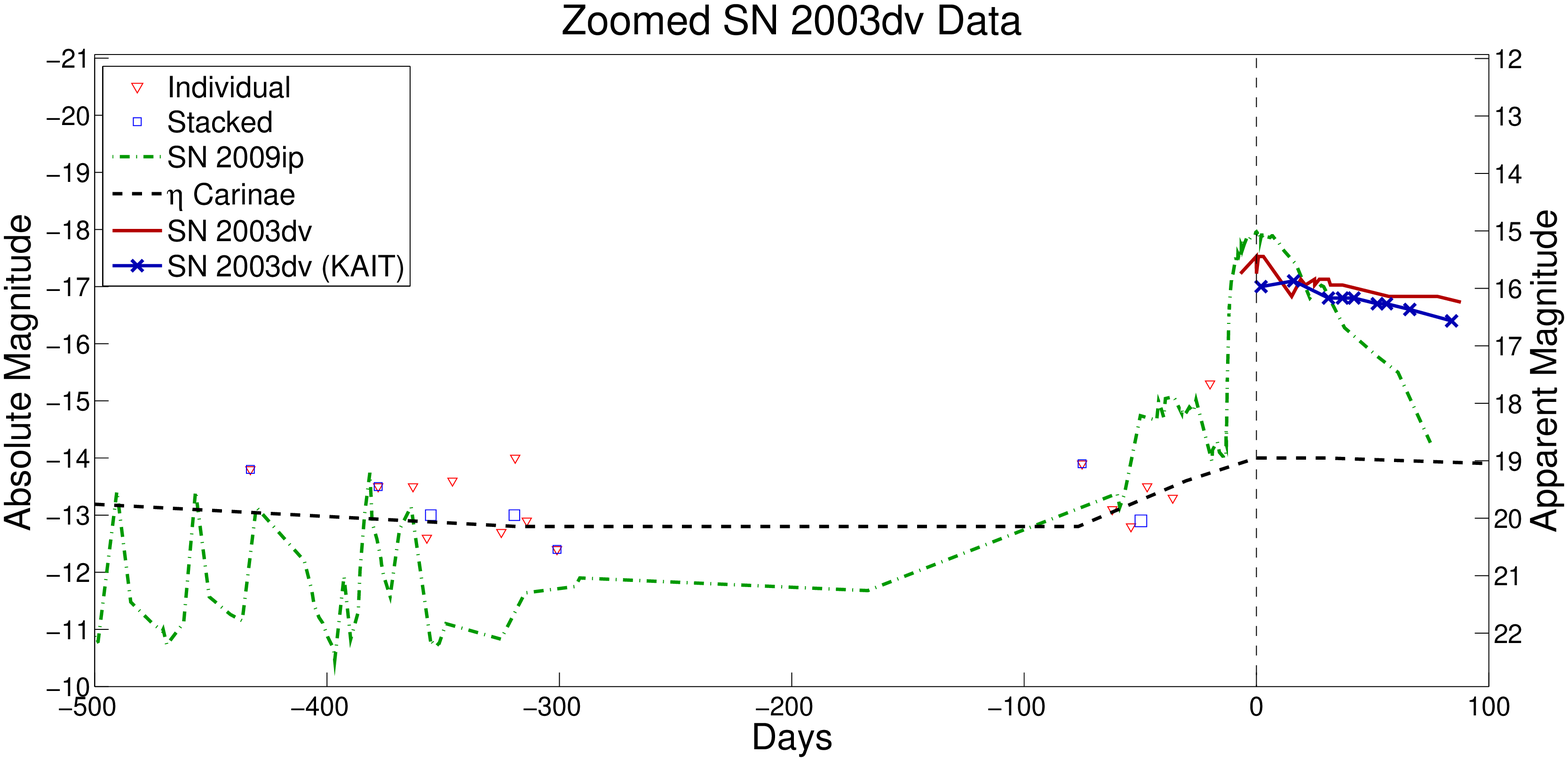}
\caption{Same as Fig. \ref{fig:LM1999el}, but for SN 2003dv.  The top plot shows all of the archival data while the bottom plot shows only data near peak magnitude.  Near-peak light-curve values for SN 2003dv were obtained from the amateur astronomy data available on the Rochester Astronomy website of nearby SNe (\protect\url{http://www.rochesterastronomy.org/sn2003/sn2003dv.html}).  We determine the date of the peak observed magnitude from this well-sampled light curve.}
\label{fig:LM2003dv}
\end{figure*}

Results for SN 2003dv are shown in Figure \ref{fig:LM2003dv} and Table \ref{tab:LM2003dv}.  SN 2003dv's dataset includes five years of limiting magnitudes, but only a handful of good images were acquired for this target each year.  Most of the limiting magnitudes fall in the range of $-13$ to $-14$ in absolute magnitude for SN 2003dv.  SN 2003dv's light curve reaches a similar peak luminosity but shows a much slower decline compared to that of SN 2009ip.

\begin{table*}
\begin{minipage}{200mm}
\caption{Limiting Magnitudes (LM) for SN 2003dv\tablenotemark{a}}
\label{tab:LM2003dv}
\hspace*{-30pt}
\begin{tabular}{ccccccc}
  \hline
	Year & Month & Day & Artificial Star LM, Individual & Artificial Star LM, Stacked & Background LM, Individual & Background LM, Stacked \\
	\hline
1999 & 1  & 28 & 18.7 & 18.7 & 19.7 & 19.7 \\
1999 & 2  & 1  & 18.6 & 19.2 & 18.9 & 19.4 \\
1999 & 2  & 27 & 19.8 & 19.2 & 19.8 & 19.4 \\
1999 & 4  & 10 & 19.7 & 19.9 & 20.0 & 20.2 \\
1999 & 4  & 15 & 20.0 & 19.9 & 19.9 & 20.2 \\
\hline
	\end{tabular}
	\hspace*{30pt}
\tablenotetext{a}{\raggedright The entirety of this table is available electronically.  A portion is displayed here for reference.}
\end{minipage}
\end{table*}

\subsubsection{SN 2006am}
\begin{figure*}
\centering
\includegraphics[width=1\textwidth,clip=true,trim=0cm 0cm 0cm 0cm]{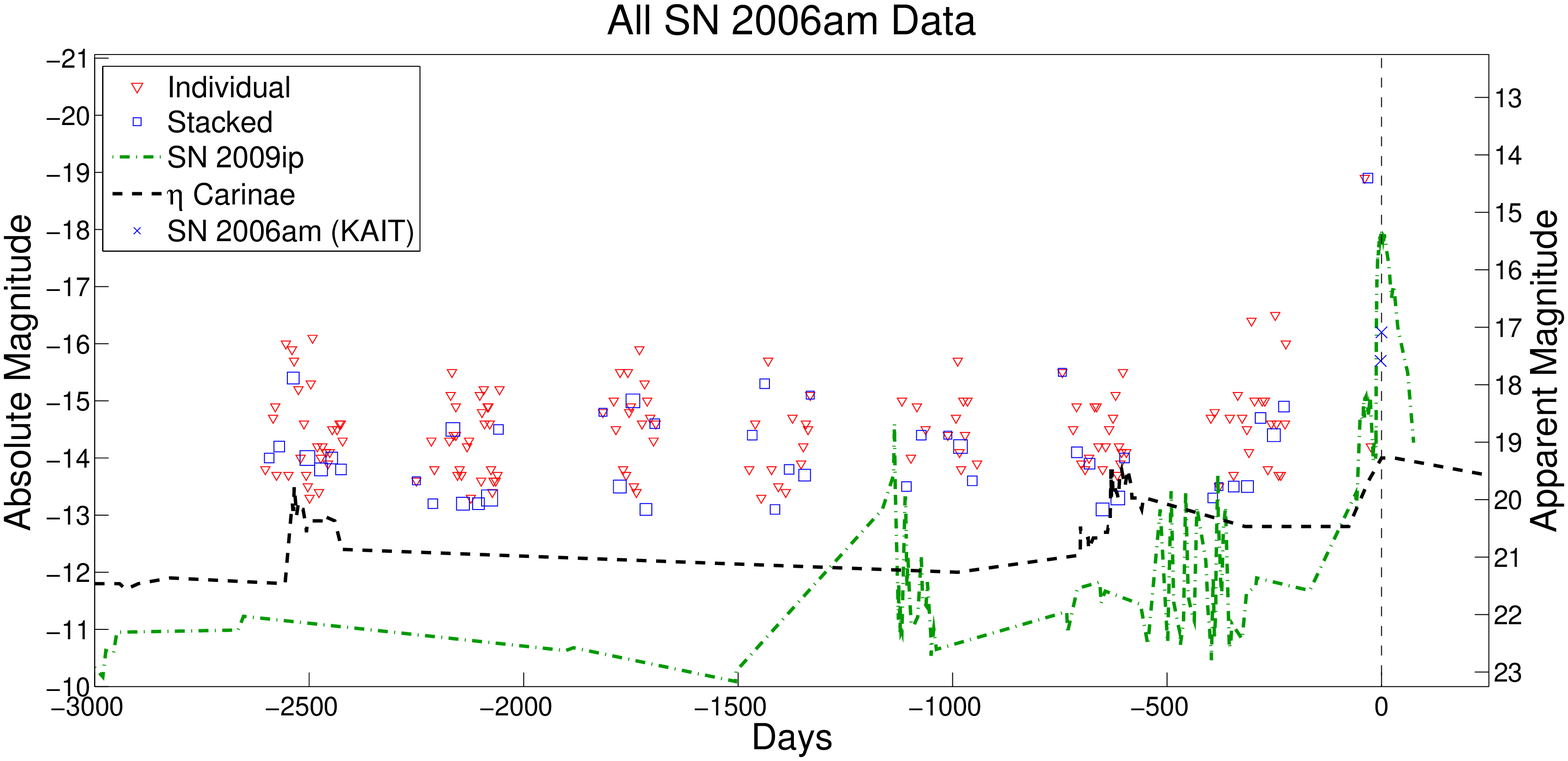} \\
\includegraphics[width=1\textwidth,clip=true,trim=0cm 0cm 0cm 0cm]{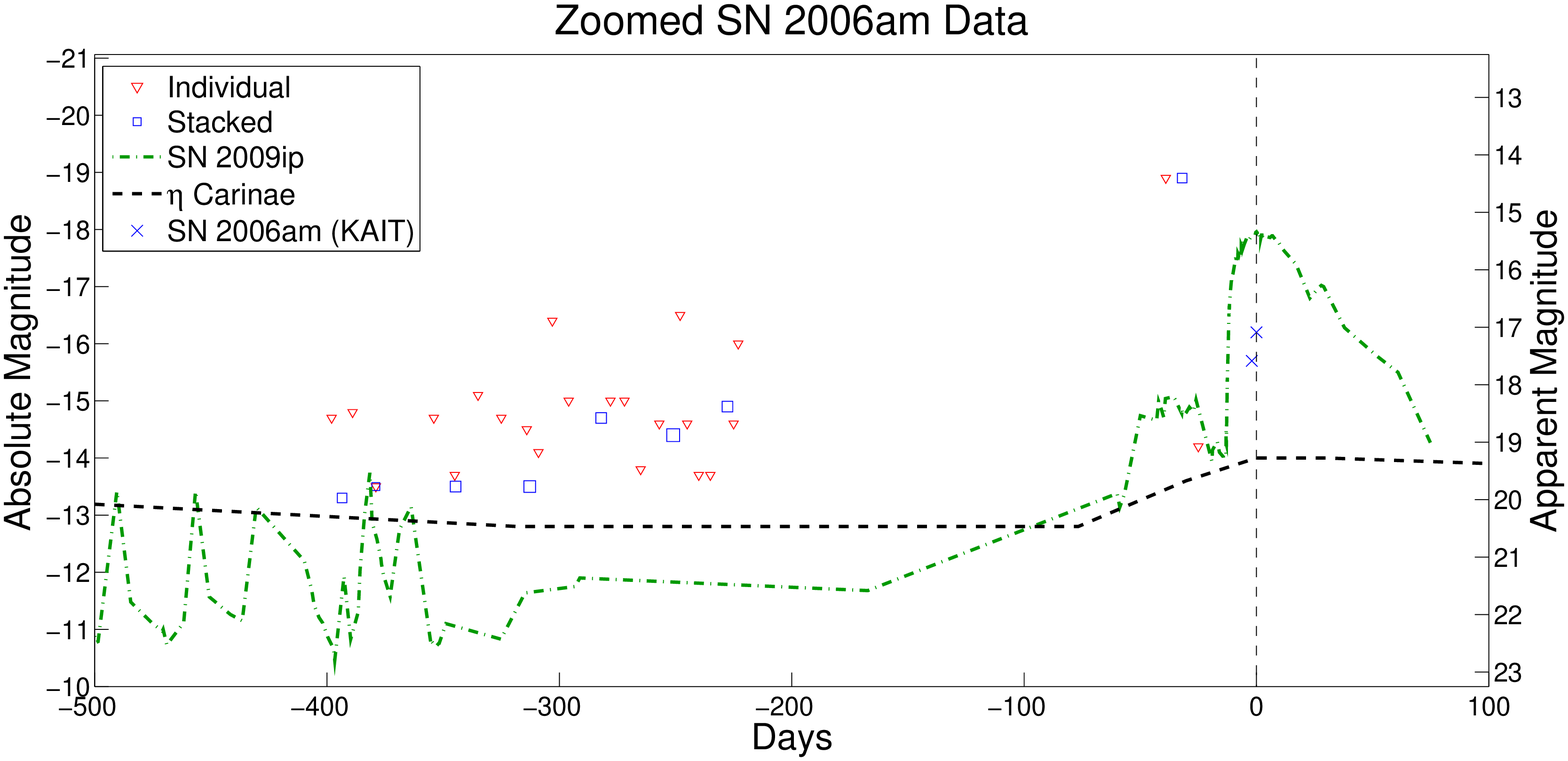}
\caption{Same as Fig. \ref{fig:LM2003dv}, but for SN 2006am.  The two near-discovery light curve values for SN 2006am were obtained from our KAIT/LOSS survey data.  We determine the date of the peak observed magnitude from this poorly sampled light curve.}
\label{fig:LM2006am}
\end{figure*}

Results for SN 2006am are shown in Figure \ref{fig:LM2006am} and Table \ref{tab:LM2006am}.  SN 2006am's dataset covers many years, but it contains only two images within the months prior to the SN explosion.  Because of its relatively faint ($M_R = -14.8$\,mag) discovery magnitude according to initial rough photometry \citep{2006IAUC.8680....1L}, SN 2006am may be very extinguished, or perhaps a SN imposter instead of a SN~IIn.  No light curve exists for SN 2006am other than our two KAIT measurements, which are the same images used earlier to discover this event.  When performing more thorough photometry on these images, however, we find a peak magnitude of $M_R \approx -16$, suggesting that SN 2006am is more likely a SN~IIn than a SN imposter.  Most of the limiting absolute magnitudes for SN 2006am lie in the range from $-13.5$ to $-15$.

\begin{table*}
\begin{minipage}{200mm}
\caption{Limiting Magnitudes (LM) for SN 2006am\tablenotemark{a}}
\label{tab:LM2006am}
\hspace*{-30pt}
\begin{tabular}{ccccccc}
  \hline
	Year & Month & Day & Artificial Star LM, Individual & Artificial Star LM, Stacked & Background LM, Individual & Background LM, Stacked \\
	\hline
1999 & 1  & 10 & 19.5 & 19.2 & 19.3 & 19.6 \\
1999 & 1  & 28 & 18.5 & 19.2 & 19.6 & 19.6 \\
1999 & 2  & 2  & 18.3 & 19.1 & 19.5 & 20.0 \\
1999 & 2  & 5  & 19.6 & 19.1 & 19.4 & 20.0 \\
1999 & 2  & 27 & 17.2 & 19.1 & 19.5 & 20.0 \\
\hline
	\end{tabular}
	\hspace*{30pt}
\tablenotetext{a}{\raggedright The entirety of this table is available electronically.  A portion is displayed here for reference.}
\end{minipage}
\end{table*}

\subsubsection{SN 2008fq}
\begin{figure*}
\centering
\includegraphics[width=1\textwidth,clip=true,trim=0cm 0cm 0cm 0cm]{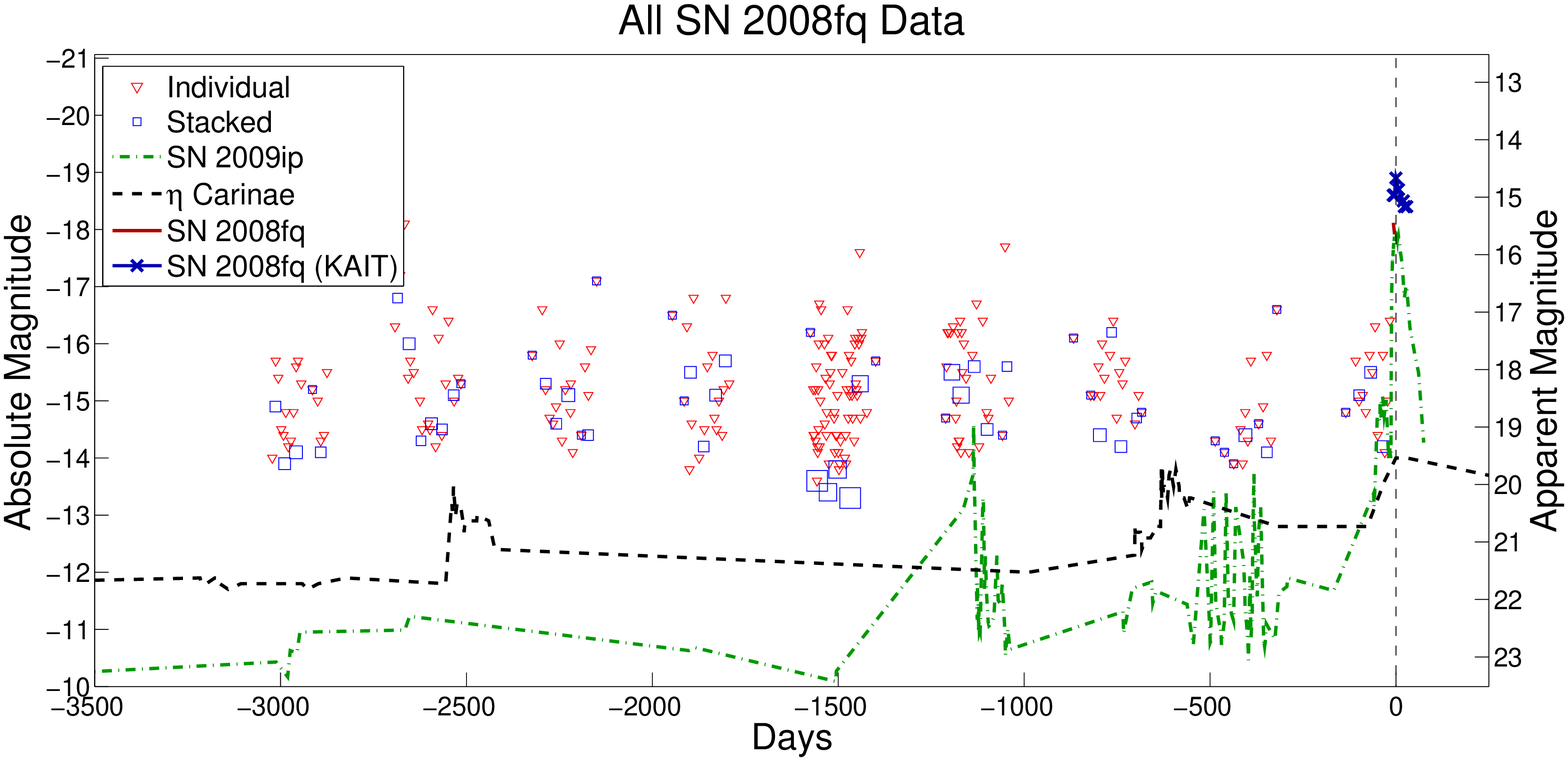} \\
\includegraphics[width=1\textwidth,clip=true,trim=0cm 0cm 0cm 0cm]{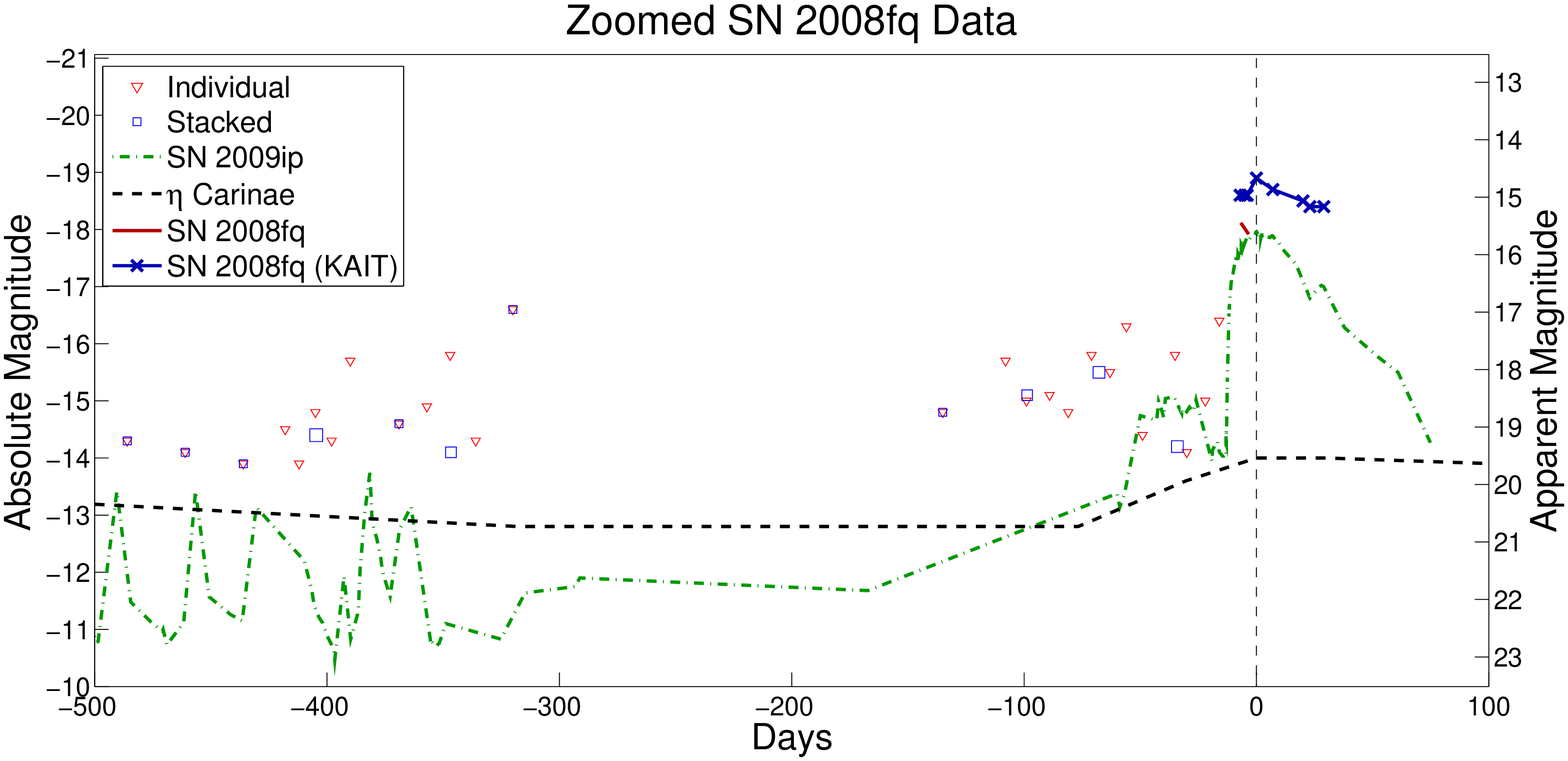}
\caption{Same as Fig. \ref{fig:LM2003dv}, but for SN 2008fq. Near-peak light-curve values for SN 2008fq were obtained from our KAIT/LOSS survey data.  Since only two data points exist on the light curve available from the Rochester Astronomy website of nearby SNe, we determine the date of the peak observed magnitude from our well-sampled KAIT light curve.}
\label{fig:LM2008fq}
\end{figure*}

Results for SN 2008fq are shown in Figure \ref{fig:LM2008fq} and Table \ref{tab:LM2008fq}.  Unfortunately, SN 2008fq is located near the brightest part of its host galaxy (NGC 6907) and at a relatively large distance of 47.4 Mpc.  This results in lower S/N and brighter limiting magnitudes for the SN 2008fq dataset.  However, the large number of images acquired for NGC 6907 mean that we have very good temporal extent and coverage for this SN, enabling us to rule out precursor outbursts in the $-14$ to $-16$ absolute magnitude range rather effectively for this SN in the years observed by KAIT. The light curve of SN 2008fq shows a slower decline than that of SN 2009ip.

\begin{table*}
\begin{minipage}{200mm}
\caption{Limiting Magnitudes (LM) for SN 2008fq\tablenotemark{a}}
\label{tab:LM2008fq}
\hspace*{-30pt}
\begin{tabular}{ccccccc}
  \hline
	Year & Month & Day & Artificial Star LM, Individual & Artificial Star LM, Stacked & Background LM, Individual & Background LM, Stacked \\
	\hline
2000 & 6  & 14 & 19.6 & 18.6 & 19.5 & 19.8 \\
2000 & 6  & 23 & 17.8 & 18.6 & 19.4 & 19.8 \\
2000 & 6  & 30 & 18.2 & 18.6 & 19.6 & 19.8 \\
2000 & 7  & 8  & 19.0 & 19.6 & 19.5 & 19.7 \\
2000 & 7  & 14 & 19.1 & 19.6 & 19.0 & 19.7 \\
\hline
	\end{tabular}
	\hspace*{30pt}
\tablenotetext{a}{\raggedright The entirety of this table is available electronically.  A portion is displayed here for reference.}
\end{minipage}
\end{table*}

\subsubsection{SN 2010jl}
\begin{figure*}
\centering
\includegraphics[width=1\textwidth,clip=true,trim=0cm 0cm 0cm 0cm]{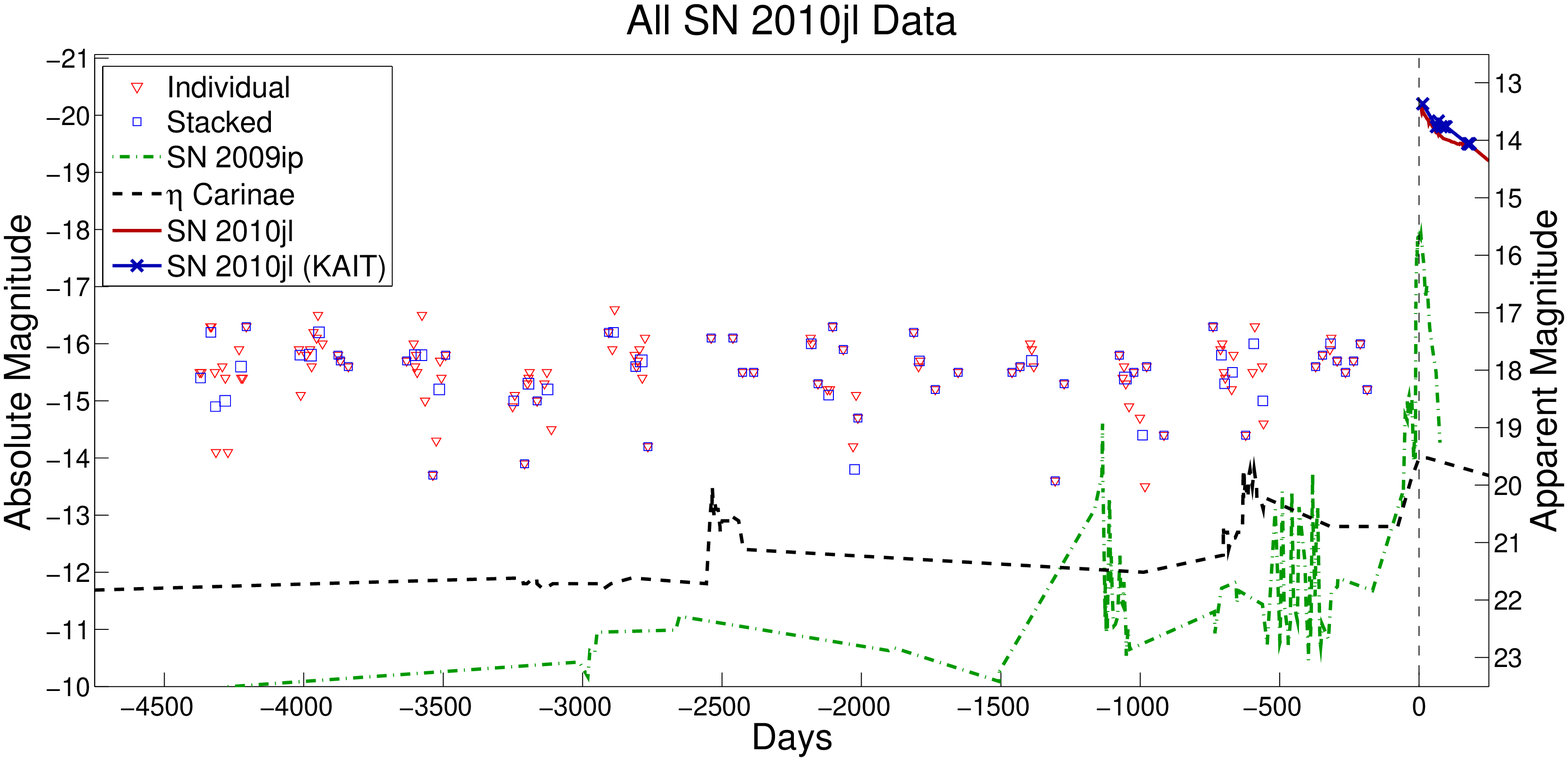} \\
\includegraphics[width=1\textwidth,clip=true,trim=0cm 0cm 0cm 0cm]{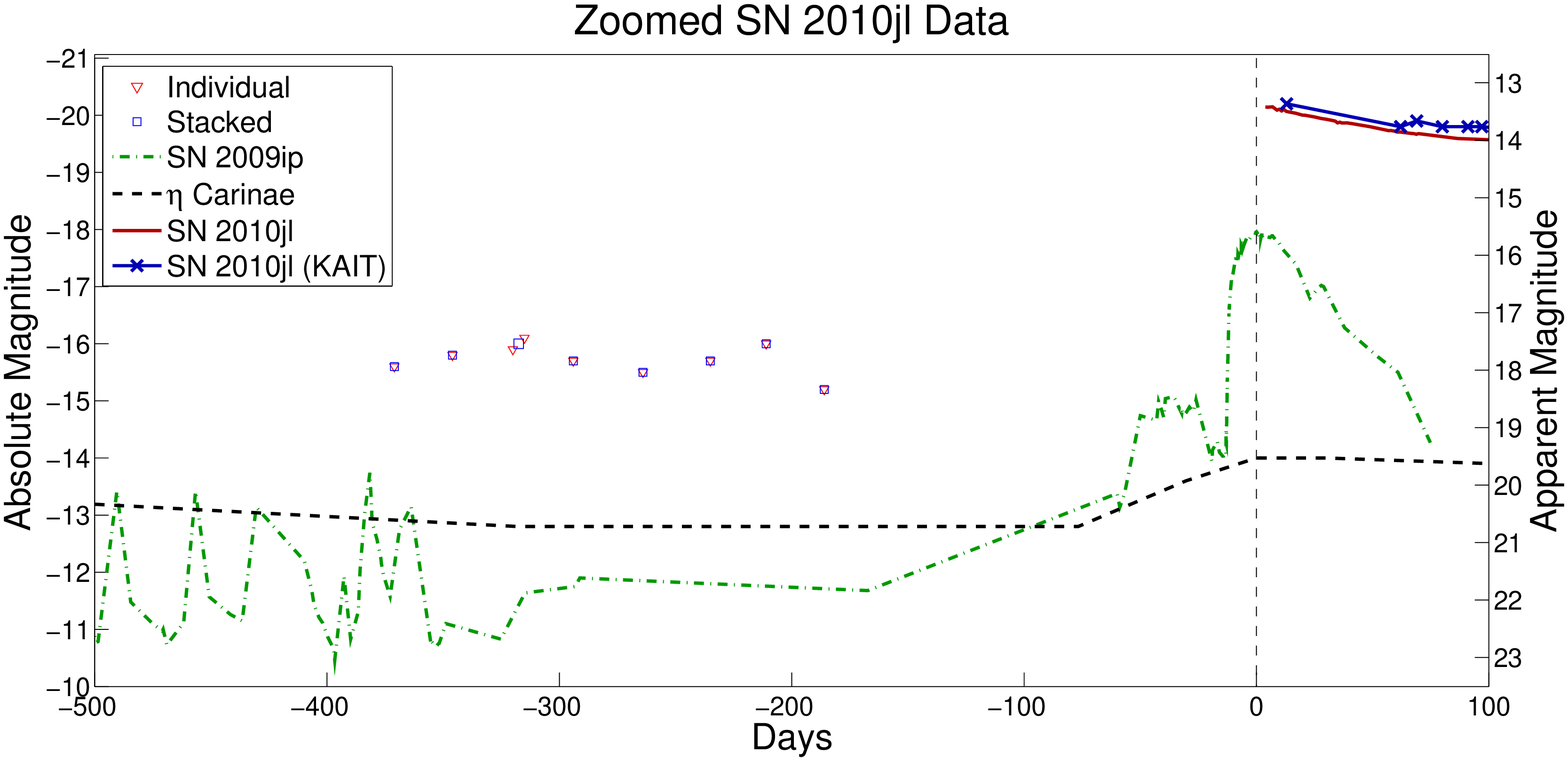}
\caption{Same as Fig. \ref{fig:LM2003dv}, but for SN 2010jl. Near-peak light-curve values for SN 2010jl were obtained from \citet{2012AJ....144..131Z}.  We determine the date of the peak observed magnitude from a light curve available on the Rochester Astronomy website of nearby SNe which extends just a few days before the published light curve.}
\label{fig:LM2010jl}
\end{figure*}

Results for SN 2010jl are shown in Figure \ref{fig:LM2010jl} and Table \ref{tab:LM2010jl}.  SN 2010jl is located in a bright part of its host galaxy (UGC 5189A), and it is also the farthest SN~IIn we consider ($d = 48.9$\,Mpc).  The number of good KAIT images of this object acquired each year is small, making this dataset noisy and sparse, even though it does cover twelve years.  With most of the limiting absolute magnitudes falling in the range of $-15$ to $-16$ mag, and none being set in the months prior to SN 2010jl's peak brightness, this dataset does not allow us to place strong constraints on the nature of SN 2010jl's progenitor.  A candidate progenitor with $M=-12.0\,\mathrm{mag}$ was, however, detected for SN 2010jl in the F300W filter on {\it HST}/WFPC2 roughly 10\,yr before peak brightness \citep{2011ApJ...732...63S}.  The light curve of SN 2010jl \citep{2012AJ....144..131Z} shows a significantly higher peak brightness than that of SN 2009ip and also a much slower decline from peak.

\begin{table*}
\begin{minipage}{200mm}
\caption{Limiting Magnitudes (LM) for SN 2010jl\tablenotemark{a}}
\label{tab:LM2010jl}
\hspace*{-30pt}
\begin{tabular}{ccccccc}
  \hline
	Year & Month & Day & Artificial Star LM, Individual & Artificial Star LM, Stacked & Background LM, Individual & Background LM, Stacked \\
	\hline
1998 & 11 & 15 & 18.0 & 18.1 & 19.4 & 19.8 \\
1998 & 11 & 21 & 18.0 & 18.1 & 19.6 & 19.8 \\
1998 & 12 & 23 & 17.2 & 17.3 & 19.8 & 20.0 \\
1998 & 12 & 27 & 17.2 & 17.3 & 19.5 & 20.0 \\
1999 & 1  & 8  & 18.0 & 18.6 & 19.6 & 19.9 \\
\hline
	\end{tabular}
\hspace*{30pt}
\tablenotetext{a}{\raggedright The entirety of this table is available electronically.  A portion is displayed here for reference.}
\end{minipage}
\end{table*}

\subsubsection{SN 2011A}
\begin{figure*}
\centering
\includegraphics[width=1\textwidth,clip=true,trim=0cm 0cm 0cm 0cm]{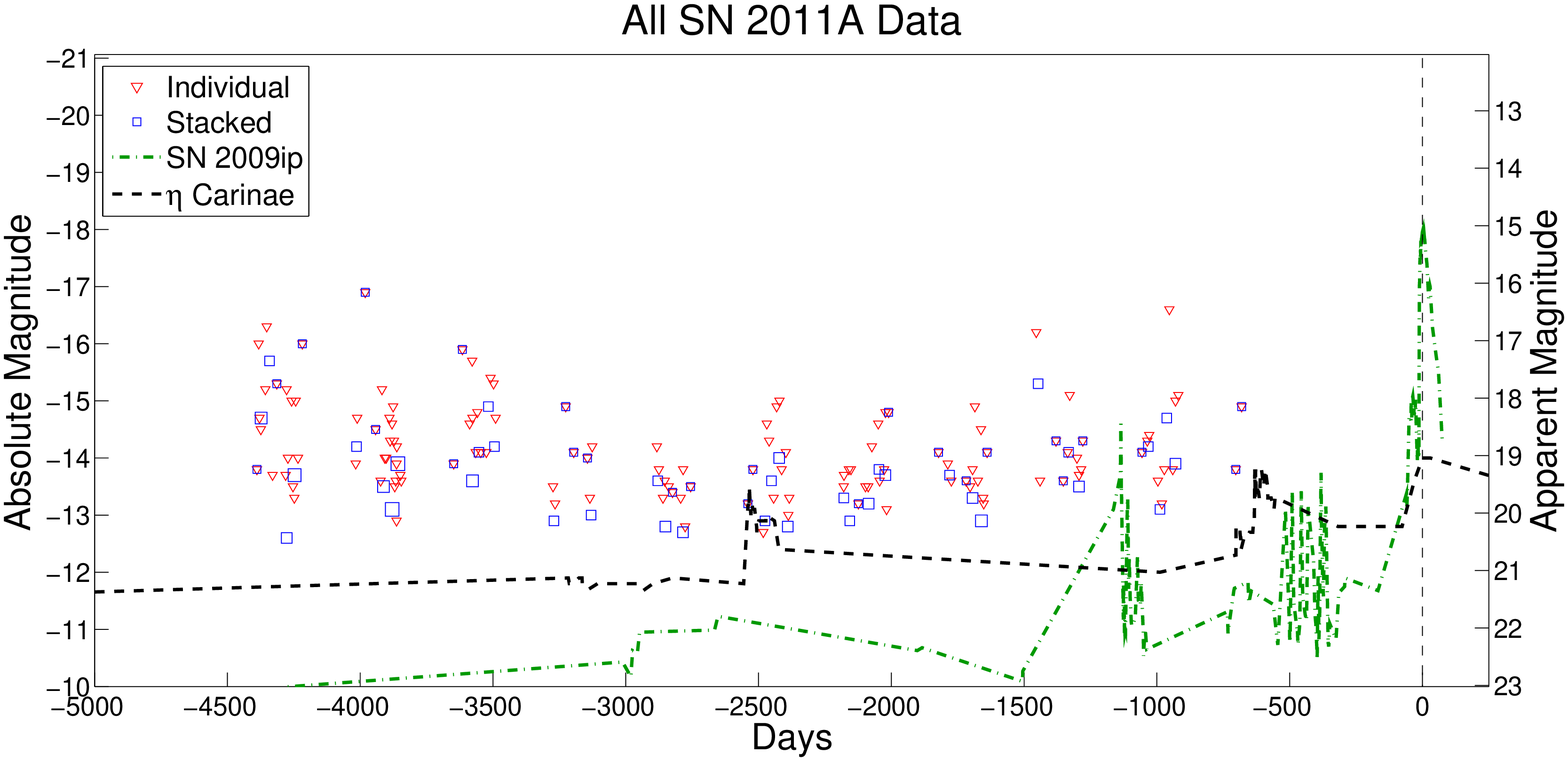}
\caption{Same as Fig. \ref{fig:LM2003dv}, but for SN 2011A.  No plot for near-peak values is presented because no survey data for SN 2011A were available for these dates.  No light curve was available for SN 2011A because of a disk failure resulting in the loss of data.}
\label{fig:LM2011A}
\end{figure*}

Results for SN 2011A are shown in Figure \ref{fig:LM2011A} and Table \ref{tab:LM2011A}.  The dataset for SN 2011A covers a large temporal extent and contains enough images each year to provide good coverage. Unfortunately, no KAIT data are available in the 1.5\,yr immediately preceding the SN explosion because of a disk failure resulting in the loss of data between September 2009 and September 2011.  Most of the limiting absolute magnitudes for SN 2011A lie in the range $-13$ to $-14$.  No light curve is available for SN 2011A to be used in comparison to SN 2009ip's light curve.

\begin{table*}
\begin{minipage}{200mm}
\caption{Limiting Magnitudes (LM) for SN 2011A\tablenotemark{a}}
\label{tab:LM2011A}
\hspace*{-30pt}
\begin{tabular}{ccccccc}
  \hline
	Year & Month & Day & Artificial Star LM, Individual & Artificial Star LM, Stacked & Background LM, Individual & Background LM, Stacked \\
	\hline
1998 & 12 & 27 & 19.1 & 19.1 & 19.0 & 19.0 \\
1999 & 1  & 3  & 16.9 & 18.2 & 19.3 & 19.6 \\
1999 & 1  & 6  & 18.2 & 18.2 & 19.2 & 19.6 \\
1999 & 1  & 11 & 18.4 & 18.2 & 18.7 & 19.6 \\
1999 & 1  & 28 & 17.7 & 18.2 & 18.1 & 19.6 \\
\hline
	\end{tabular}
	\hspace*{30pt}
\tablenotetext{a}{\raggedright The entirety of this table is available electronically.  A portion is displayed here for reference.}
\end{minipage}
\end{table*}

\section{Discussion}
\label{sec:Dis}

In general, the limiting magnitudes that we set prior to our sample of SNe~IIn are not faint enough to rule out SN 2009ip or $\eta$ Carinae-like eruptions with high confidence.  Even the stacked images are usually not sufficiently deep because the KAIT survey does not typically acquire a large number ($>10$) of images of each particular galaxy each month.  Thus, a main conclusion is that the lack of detections of pre-SN outbursts is not very surprising, even if all SNe~IIn are core-collapse events with substantial ($M_R \approx -14$\,mag) precursor eruptions.  Nevertheless, we can perform a number of statistical tests, which we discuss below, to gain a better understanding of what our limits imply with regard to SN~IIn progenitor properties, and to guide future observing strategies.

\subsection{Coverage Rates for Example Outbursts}
\begin{figure*}
\centering
\includegraphics[width=0.8\textwidth,clip=true,trim=0cm 0cm 0cm 0cm]{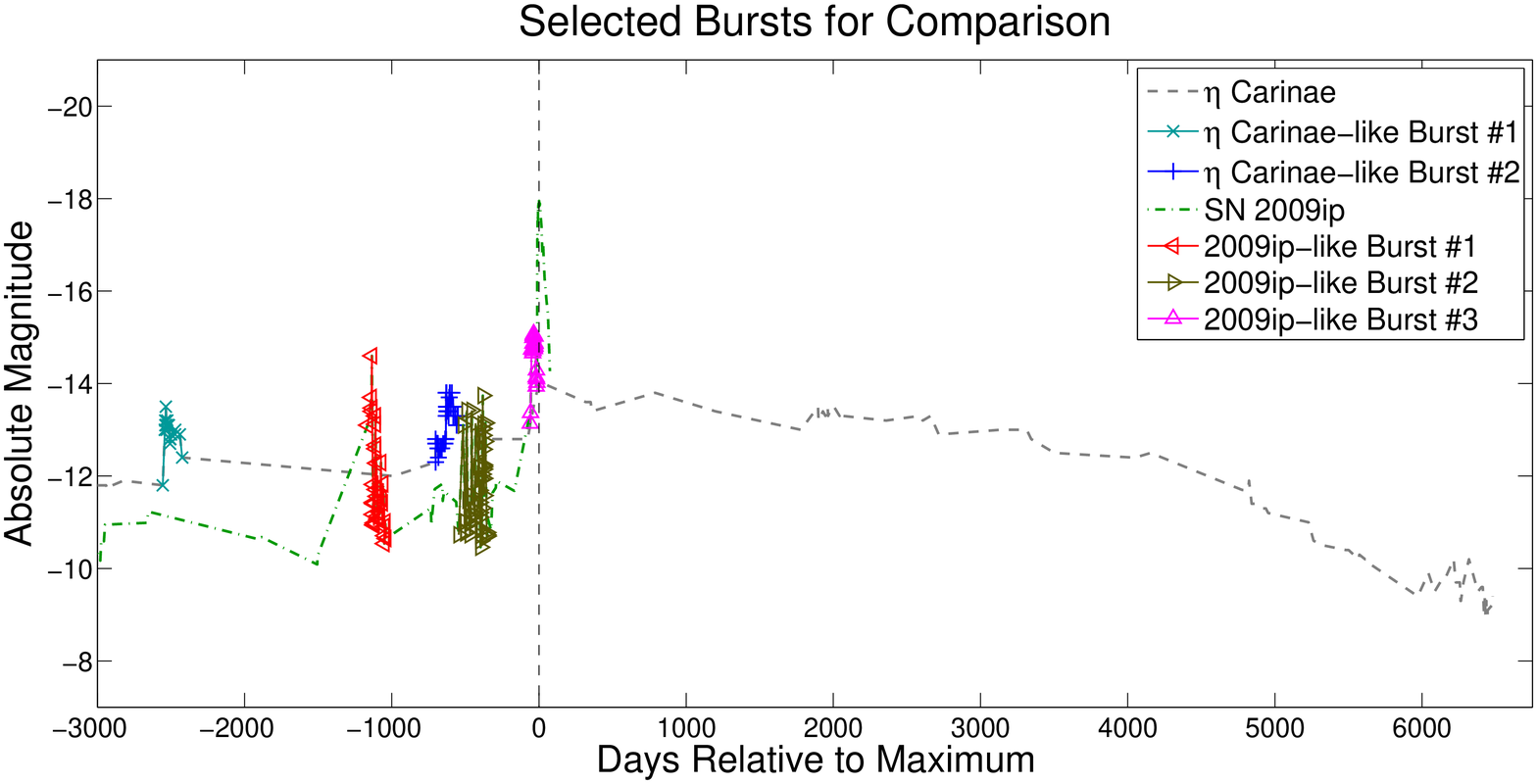}\\
\includegraphics[width=0.8\textwidth,clip=true,trim=0cm 0cm 0cm 0cm]{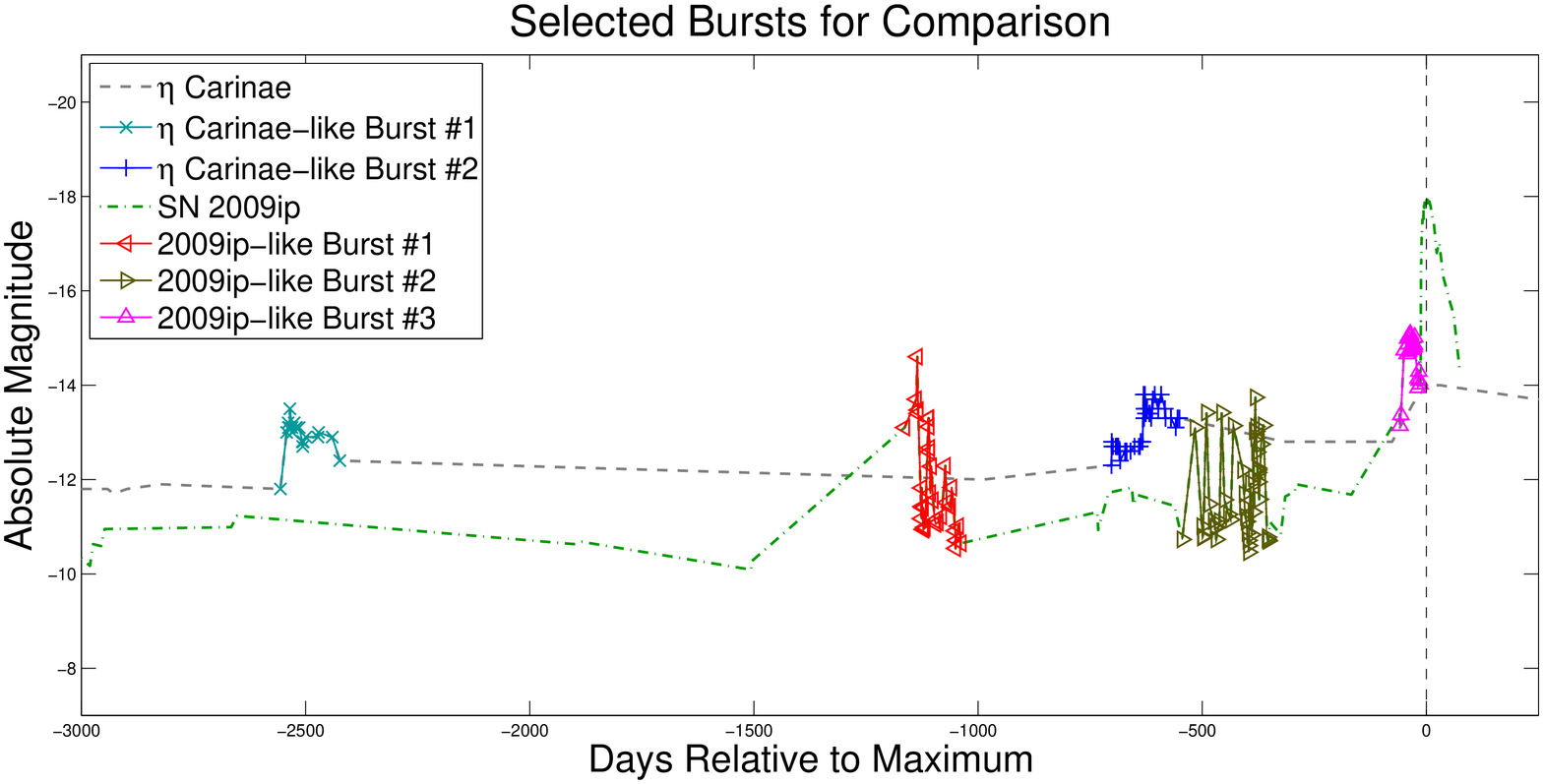}    
\caption{The top plot shows the entire light curves for SN 2009ip and $\eta$ Carinae, while the bottom is zoomed in to cover only $\sim3000$ days.  The highlighted regions show the selected outbursts used for comparison to the calculated limiting magnitudes of each SN~IIn.  SN 2009ip-like Burst \#3 is actually the SN's initial rise rather than a precursor outburst.}
\label{fig:SelOut}
\end{figure*}

Although control times were not calculated for the KAIT data for any SN imposters, we compute fractional coverage rates for five different known LBV-like outbursts, which are shown in Figure \ref{fig:SelOut}.  Three of these outbursts are taken from the SN 2009ip light curve (the last of which is actually likely to be the start of the SN explosion itself; see \citealp[]{2013MNRAS.430.1801M,2014MNRAS.438.1191S}) and the other two are taken from $\eta$ Carinae's historical light curve \citep{2011MNRAS.415.2009S}.  The length of the outbursts varies from 45 to 192 days and some of the outbursts, like that of the second SN 2009ip-like burst, include many brief individual outbursts within them.  Overall, the outbursts cover a range from $\sim-11\,\mathrm{mag}$ during a quiescent part of the burst to $\sim-15\,\mathrm{mag}$ at peak outburst, consistent with the observed distribution of SN imposter peak luminosities \citep{2011MNRAS.415..773S}. Observations of SN 2009ip were taken in either an unfiltered bandpass or the $R$ band, while observations of $\eta$ Carinae were made primarily by eye in the visible.  With the temperatures expected for these outbursts near peak ($\sim6000$--7000\,K), $R$-band and visible photometry should be comparable (because the bolometric correction is small), except in the case of very high H$\alpha$ emission levels.  

The fractional coverage rates we compute are essentially the probabilities that we would have been able to detect each particular outburst.  We assume each of the outbursts has an equal likelihood of having occurred on any day since our first available observation for each SN.  If the outburst is ever brighter than a limiting magnitude we have set for a given day, then we claim that we would have been able to detect that outburst.  After inserting each outburst on every possible day that it could have occurred, we divide the total number of detected outbursts by the total number of outbursts inserted to find the fractional coverage rate.  The results of this assessment are presented in Table \ref{tab:SNCoverage}.

\begin{table*}
\begin{minipage}{200mm}
\caption{Fractional Probability of Detecting Given Outbursts Using Artificial Star Injection LMs}
\label{tab:SNCoverage}
\begin{tabular}{ccccccc}
  \hline
	SN & Image Type & 2009ip Burst \#1 & 2009ip Burst \#2 & 2009ip Burst \#3 & $\eta$ Car Burst \#1 & $\eta$ Car Burst \#2 \\
	\hline
	1999el	&	Individual	&	0.48	&	0.54	&	0.56	&	0.53	&	0.62	\\
-	&	Stacked	&	0.20	&	0.17	&	0.46	&	0.34	&	0.40	\\
2003dv	&	Individual	&	0.32	&	0.18	&	0.48	&	0.31	&	0.47	\\
-	&	Stacked	&	0.23	&	0.12	&	0.43	&	0.25	&	0.45	\\
2006am	&	Individual	&	0.11	&	0.01	&	0.49	&	0.01	&	0.17	\\
-	&	Stacked	&	0.13	&	0.02	&	0.43	&	0.03	&	0.27	\\
2008fq	&	Individual	&	0.02	&	0.00	&	0.35	&	0.00	&	0.01	\\
-	&	Stacked	&	0.02	&	0.00	&	0.27	&	0.00	&	0.04	\\
2010jl	&	Individual	&	0.01	&	0.00	&	0.11	&	0.00	&	0.02	\\
-	&	Stacked	&	0.00	&	0.00	&	0.07	&	0.00	&	0.01	\\
2011A	&	Individual	&	0.15	&	0.04	&	0.40	&	0.08	&	0.26	\\
-	&	Stacked	&	0.14	&	0.06	&	0.36	&	0.16	&	0.26	\\
Collective\tablenotemark{a}	&	Individual	&	0.74	&	0.64	&	0.96	&	0.70	&	0.88	\\
-	&	Stacked	&	0.56	&	0.33	&	0.92	&	0.60	&	0.83 \\
	\hline
	\end{tabular}
\tablenotetext{a}{\raggedright These are the probabilities that we would have detected at least one outburst from any of the SN~IIn targets.}
\end{minipage}
\end{table*}

Because of the brightness of the host galaxies at the location of SN 2010jl and SN 2008fq, the limiting magnitudes are too bright to place a strong constraint on the potential presence of pre-SN outbursts.  However, the other four SNe have reasonable ($\ge0.10$) fractional coverage rates for most of the example outbursts, with particularly high ($\ge0.35$) coverage rates for SN 2009ip-like burst \#3 simulations.  Considering that we have no limiting magnitudes for a large portion of the year owing to right-ascension constraints, any coverage rate near 0.5 is high.

Even though we have the highest fractional coverage rates for SN 2009ip-like burst \#3, we focus more on the other outbursts in this discussion because this event (2012a) is likely to be the SN explosion itself, which is initially faint because of having a BSG progenitor \citep{2013MNRAS.430.1801M,2014MNRAS.438.1191S}, as was the case for SN 1987A.  Our next most prominent outburst is that of $\eta$ Car-like burst \#2 (1843), which shows rates above 17\% for all of our objects, excluding SN 2010jl and SN 2008fq.  If all of the SNe~IIn in our sample had gone through $\eta$ Car-like burst \#2 phases within their recent histories, then we would have an 88\% chance of detecting at least one outburst from the six targets.  Even for SN 2009ip-like burst \#2, the faintest and shortest duration outburst, we find a 64\% chance of having detected at least one outburst among all of our events.  

\subsection{Possible Extinction from Circumstellar Dust}
These calculations are heavily dependent on the fact that we ignore local extinction by circumstellar dust (though SN 2009ip and $\eta$ Carinae may have been surrounded by dust as well).  We consider the possibility of circumstellar dust enshrouding our targets by adding the effects of artificial extinction to the example outbursts and recalculating the coverage rates.  We find that with just $A_R\ge0.5\,\mathrm{mag}$, the probability that we would have detected at least one outburst from any of our targets similar to that of SN 2009ip's burst \#2 drops below 10\%.  Thus, it may be that the majority of SN~IIn precursors are somewhat less luminous intrinsically than SN 2009ip or $\eta$ Carinae, or that they appear fainter because of a modest amount of dust that formed in a previous eruption.  The dust formed by precursor eruptions may be destroyed by the SN explosion.  Because of this, we cannot use measurements of reddening along the line of sight to SNe to rule out extinction effects on the progenitor systems due to CSM dust.

\subsection{Coverage Rates for Simulated Outbursts}
\begin{figure*}
\centering
\includegraphics[width=0.40\textwidth,clip=true,trim=0cm 0cm 0cm 0cm]{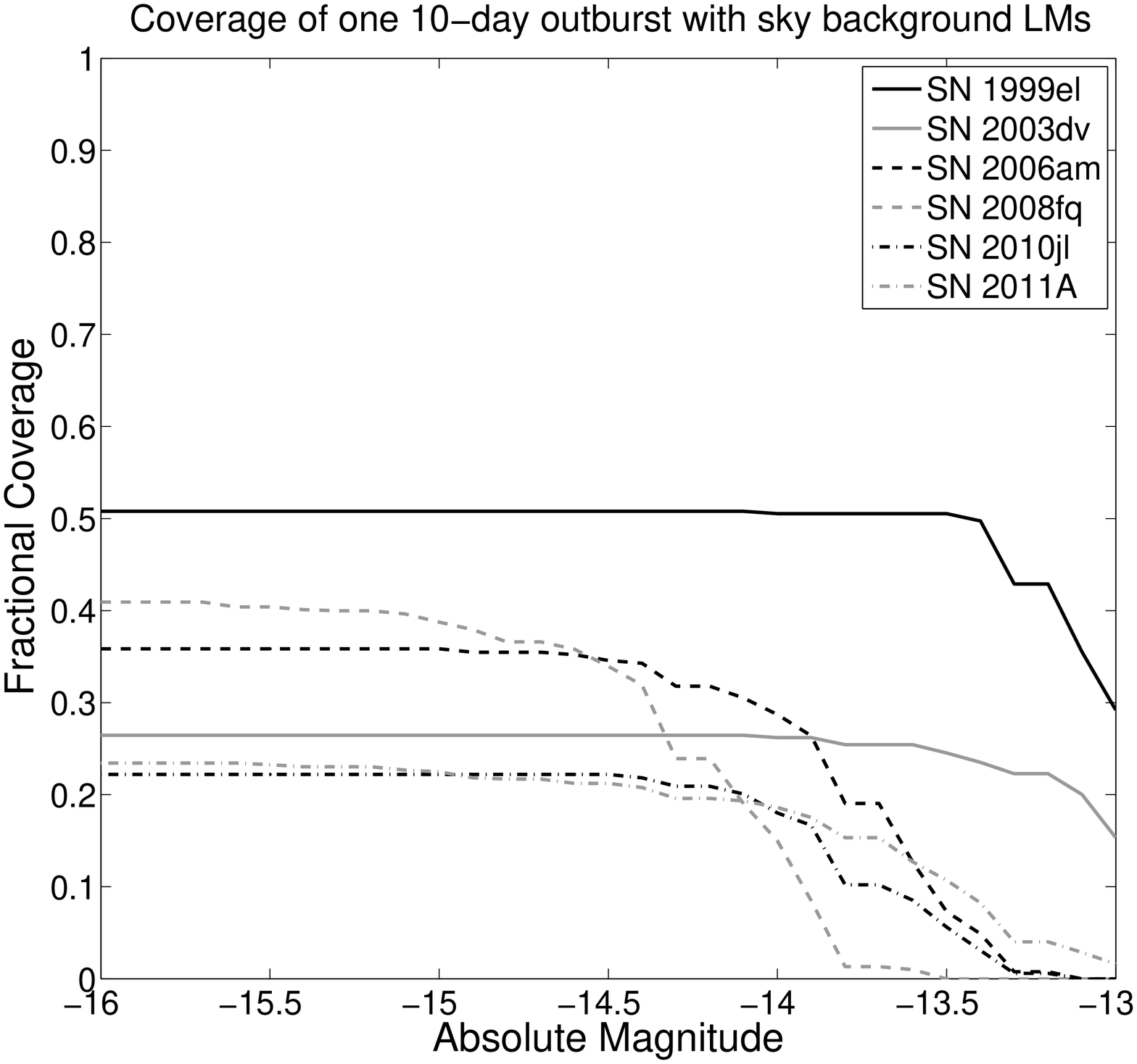} 
\includegraphics[width=0.40\textwidth,clip=true,trim=0cm 0cm 0cm 0cm]{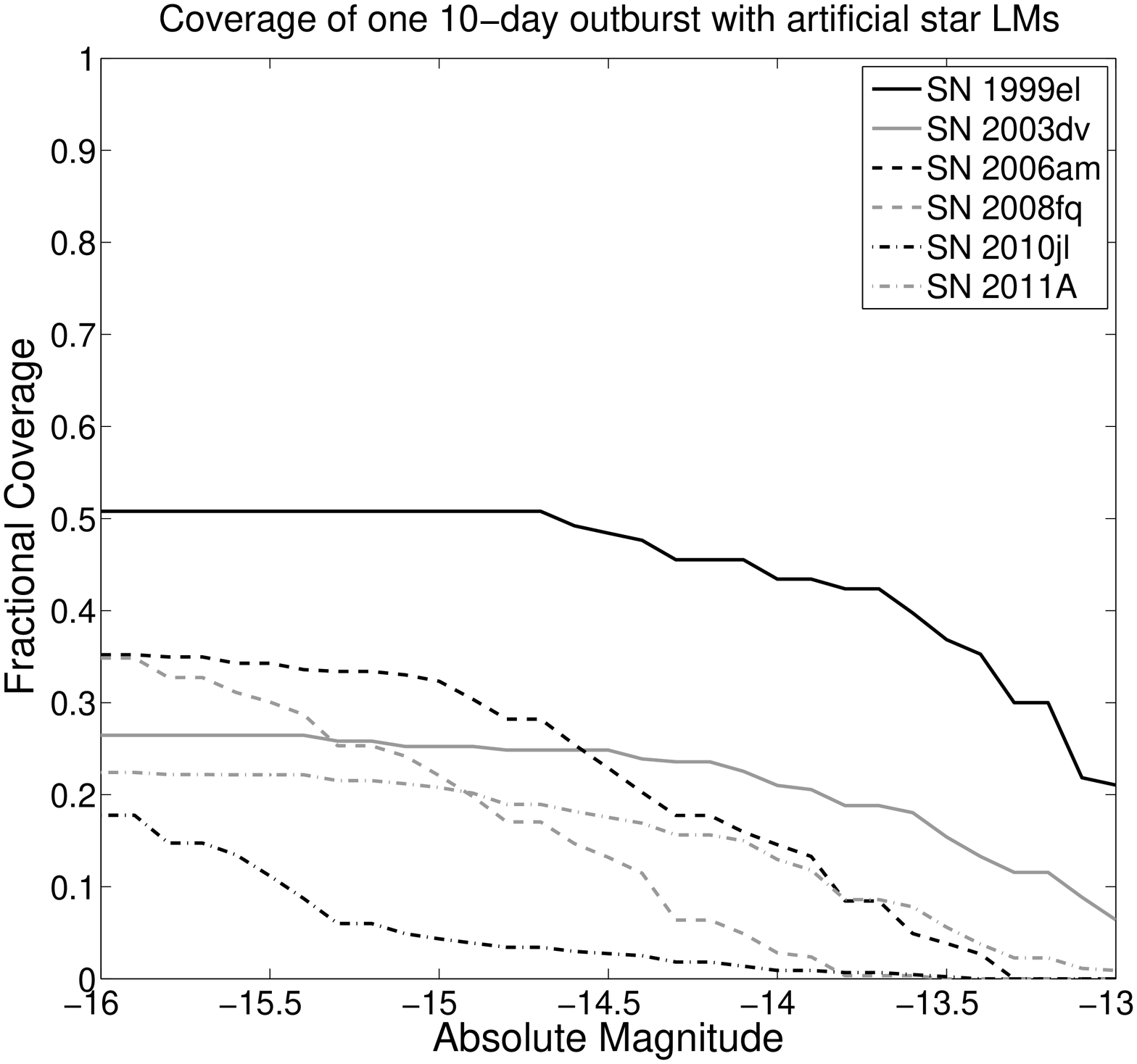} 
\includegraphics[width=0.40\textwidth,clip=true,trim=0cm 0cm 0cm 0cm]{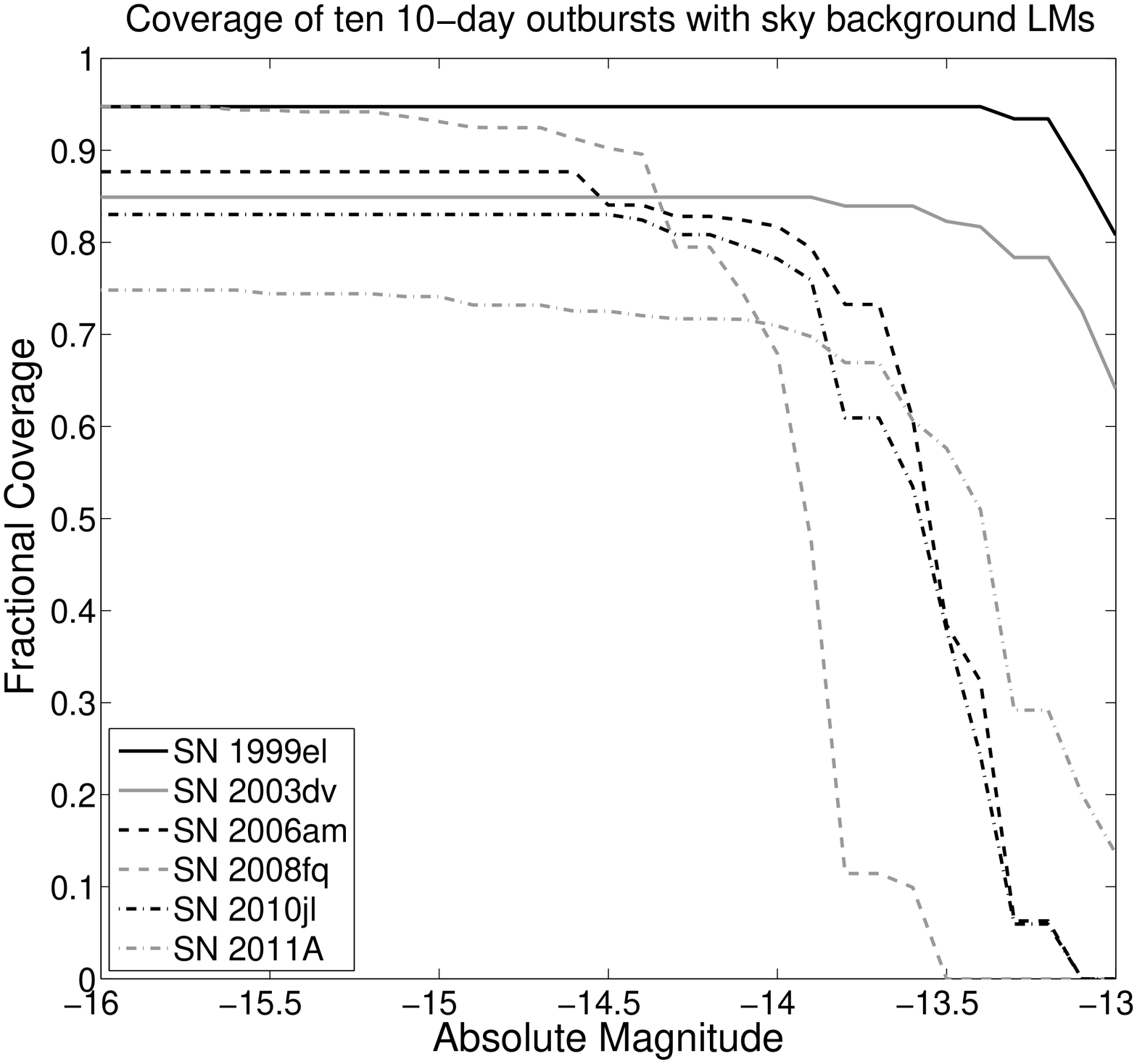} 
\includegraphics[width=0.40\textwidth,clip=true,trim=0cm 0cm 0cm 0cm]{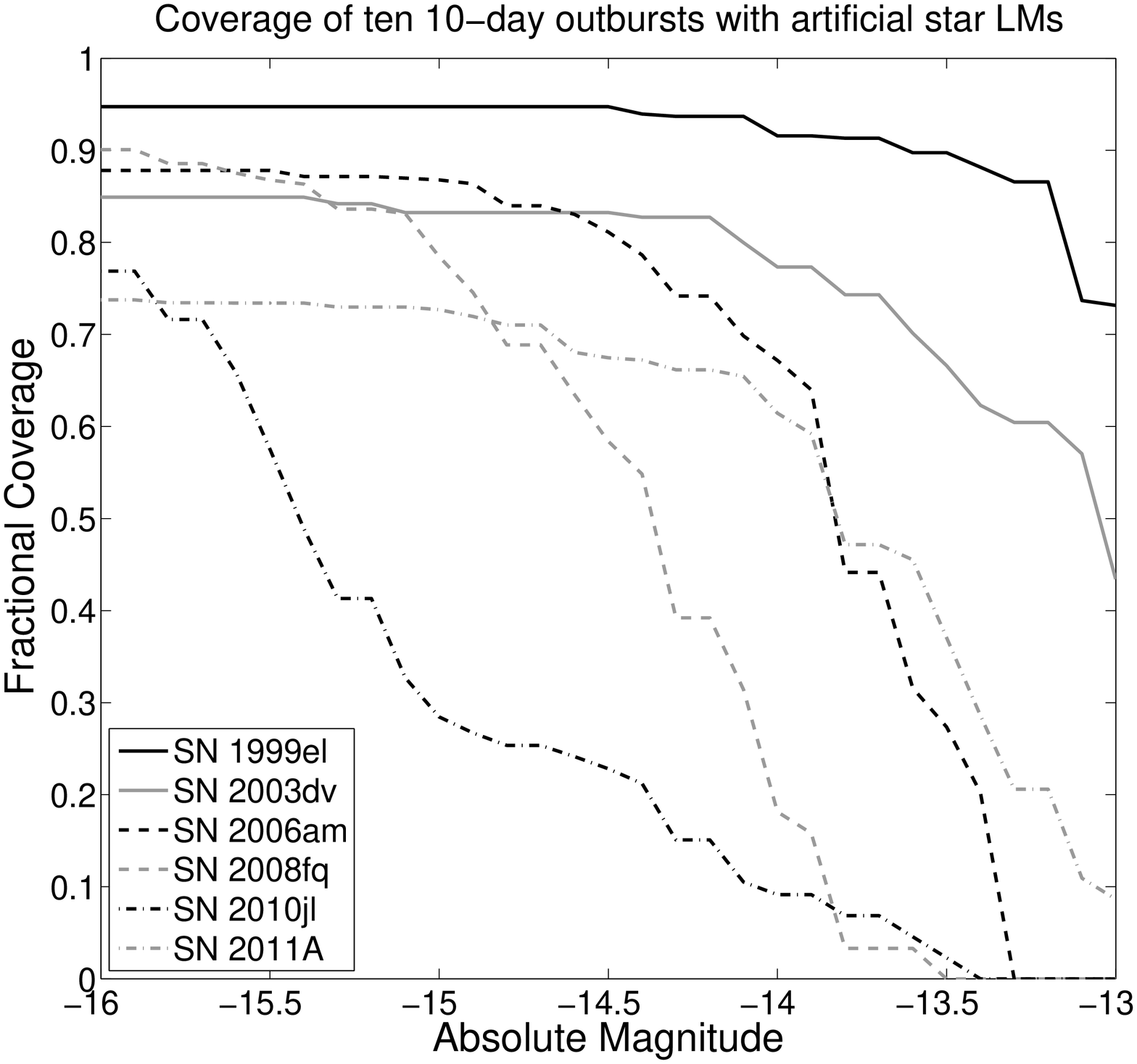}
\includegraphics[width=0.40\textwidth,clip=true,trim=0cm 0cm 0cm 0cm]{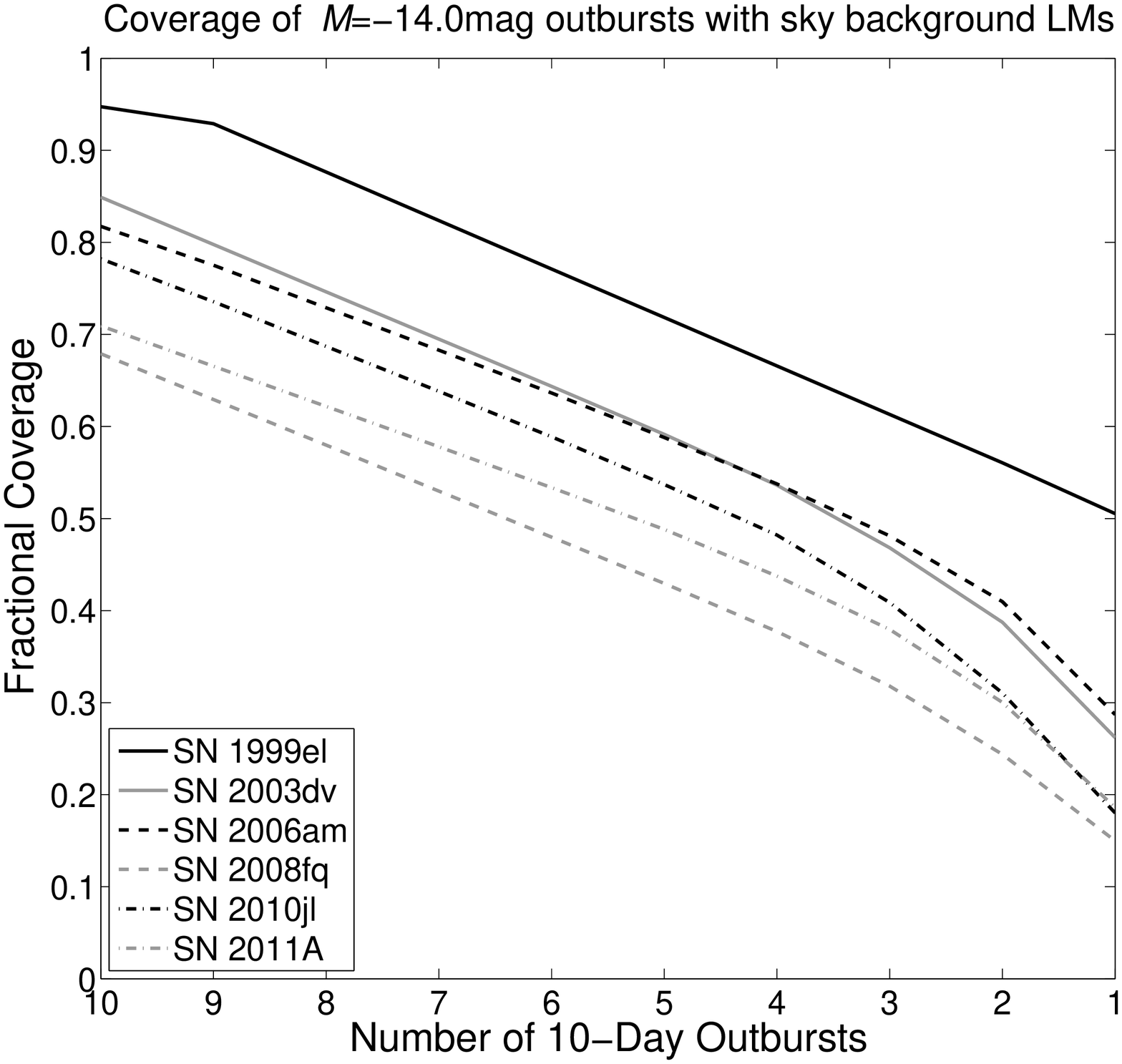} 
\includegraphics[width=0.40\textwidth,clip=true,trim=0cm 0cm 0cm 0cm]{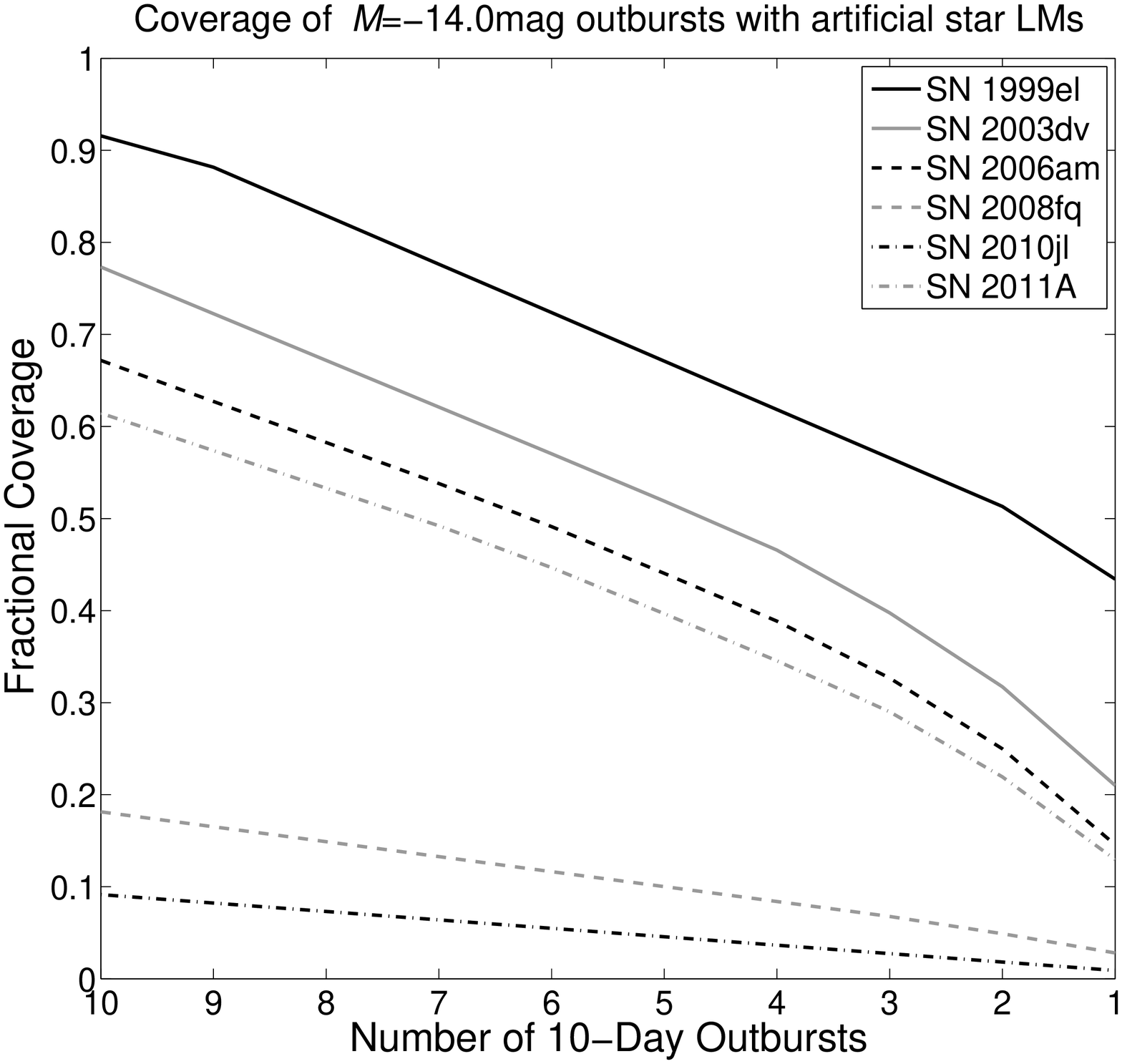}  
\caption{Fractional coverage of SN 2009ip-like (many quick) simulated outbursts based on the limiting magnitudes for the various SNe~IIn.  The left column uses the limiting magnitudes based on background noise, whereas the right column uses the limiting magnitudes based on artificial star injection.  The top row shows the fractional coverage for varying magnitudes of a 10-day simulated outburst.  The middle row provides the fractional coverage for varying magnitudes of ten 10-day simulated outbursts.  The bottom row gives the fractional coverage for varying numbers of 10-day outbursts at a fixed magnitude of $M=-14$.}
\label{fig:SimIP}
\end{figure*}

\begin{figure*}
\centering
\includegraphics[width=0.40\textwidth,clip=true,trim=0cm 0cm 0cm 0cm]{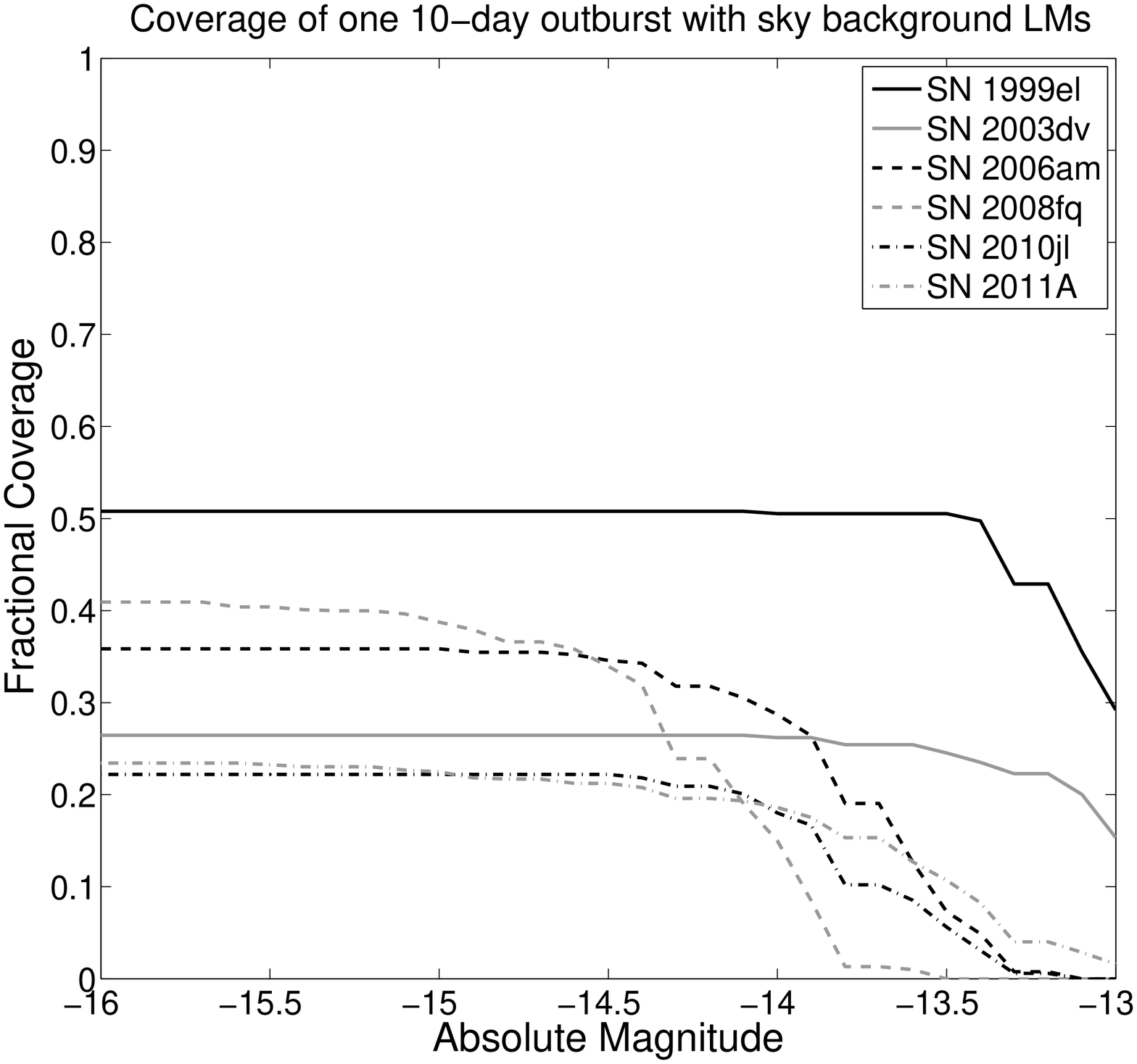}
\includegraphics[width=0.40\textwidth,clip=true,trim=0cm 0cm 0cm 0cm]{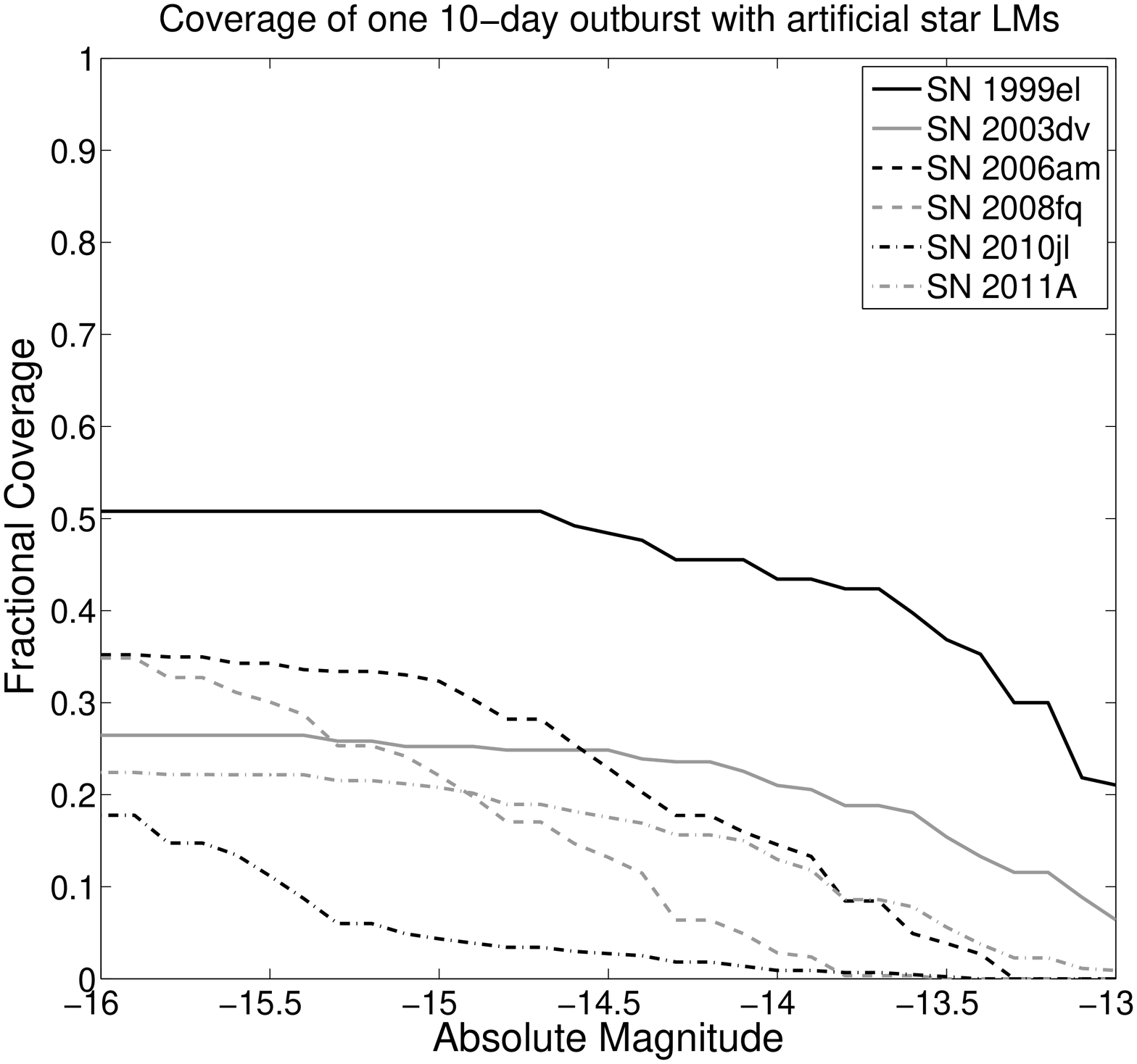} 
\includegraphics[width=0.40\textwidth,clip=true,trim=0cm 0cm 0cm 0cm]{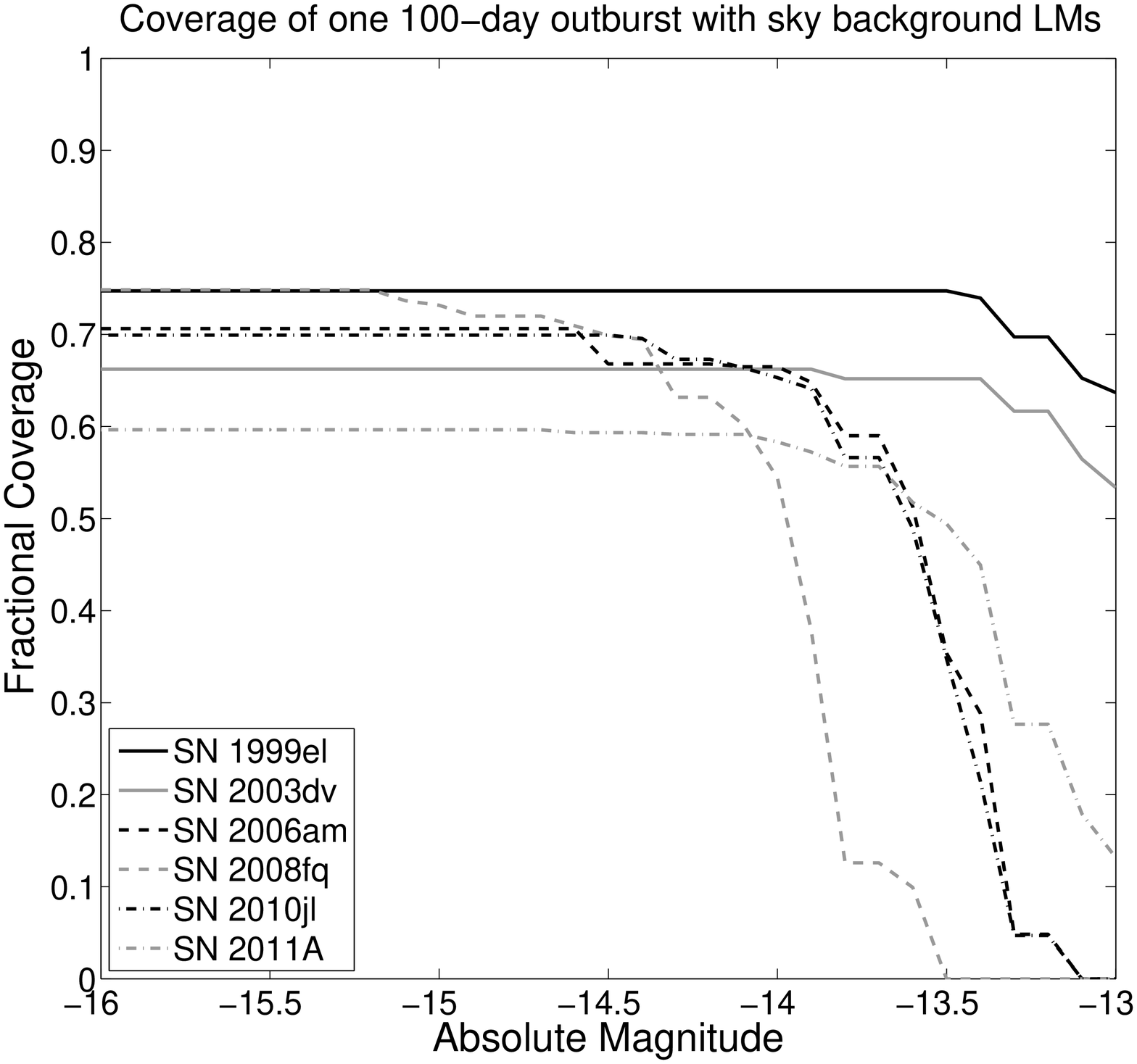} 
\includegraphics[width=0.40\textwidth,clip=true,trim=0cm 0cm 0cm 0cm]{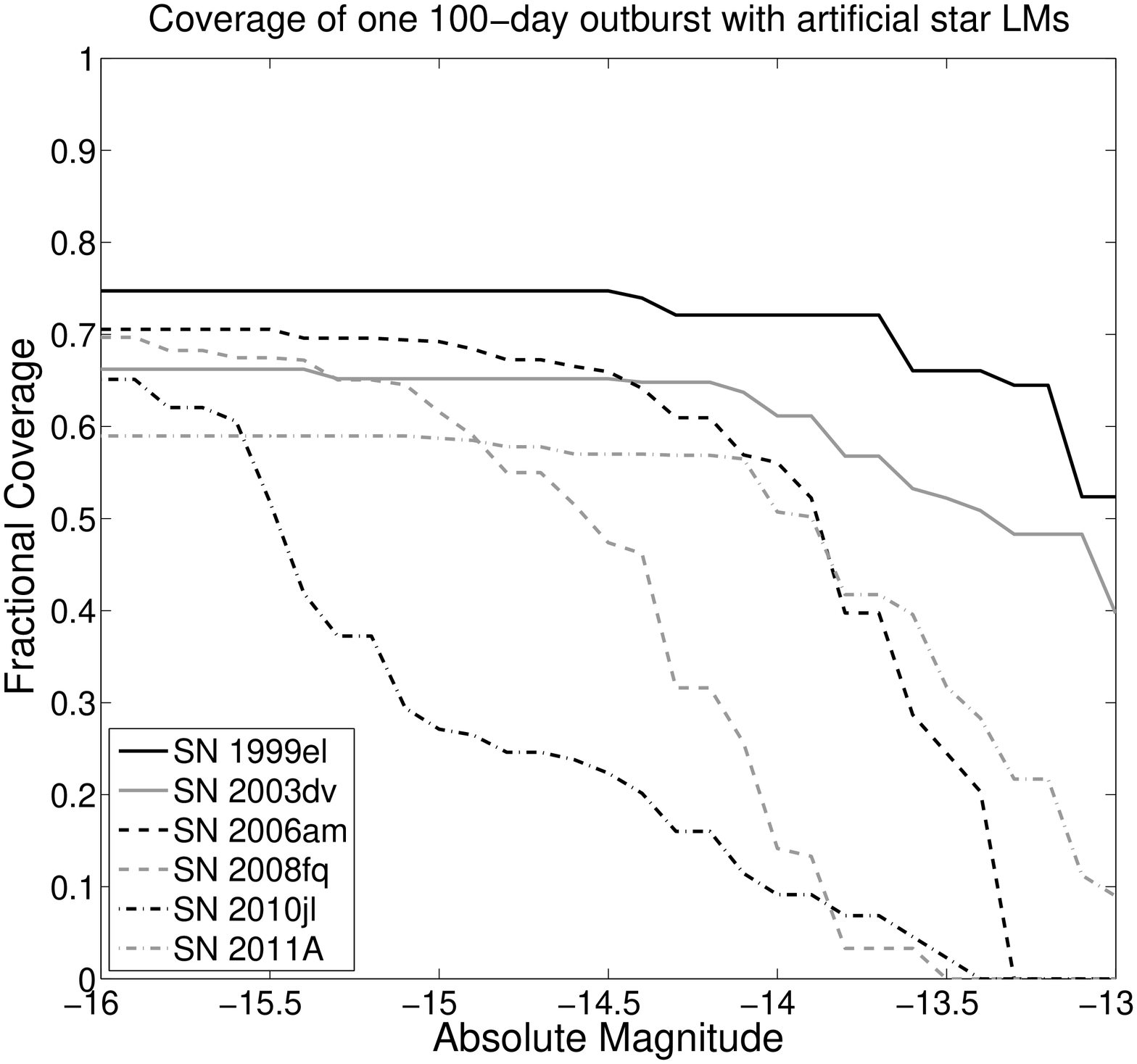} 
\includegraphics[width=0.40\textwidth,clip=true,trim=0cm 0cm 0cm 0cm]{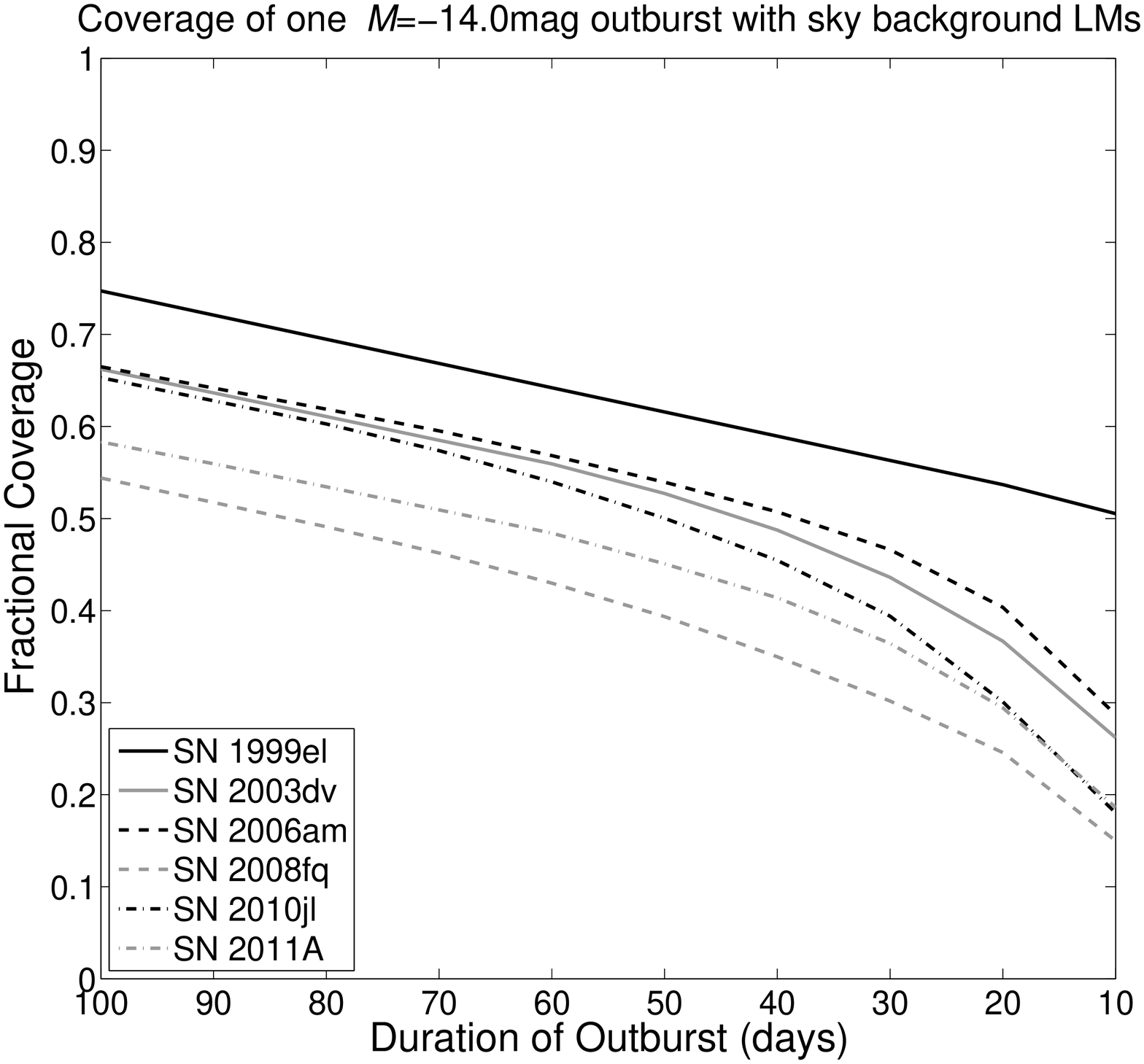} 
\includegraphics[width=0.40\textwidth,clip=true,trim=0cm 0cm 0cm 0cm]{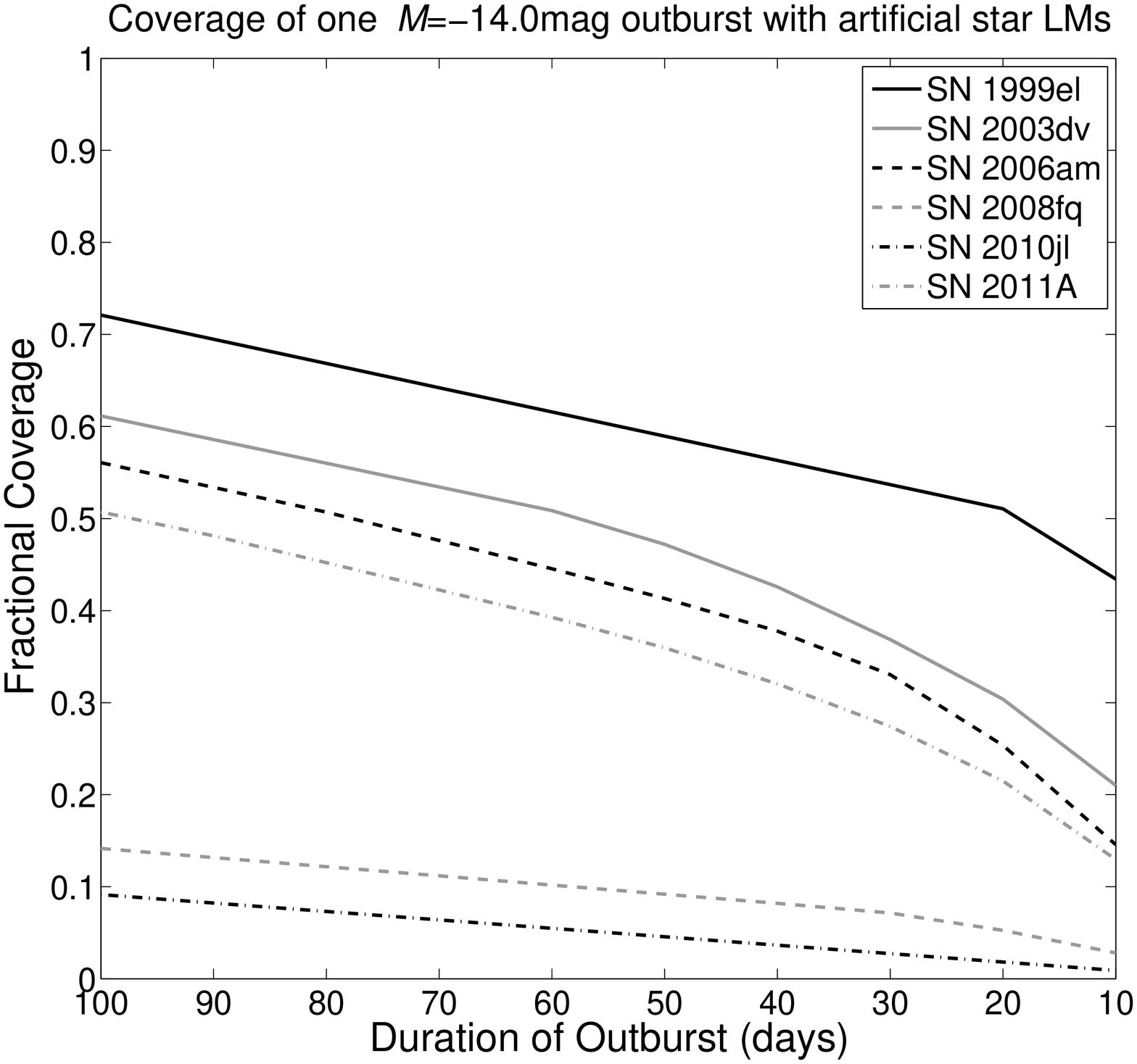} 
\caption{Same as Fig. \ref{fig:SimIP}, except for $\eta$ Carinae-like simulated outbursts. The middle row shows the fractional coverage for varying magnitudes of a single 100-day simulated outburst.  The bottom row gives the fractional coverage for varying durations of a single outburst at a fixed magnitude of $M=-14$.}
\label{fig:SimEta}
\end{figure*}

In order to further constrain the types of outbursts, we compute fractional coverage rates (the chance that we would have detected a given outburst if it had occurred on any random day since the start of our observations for the SN in question) for simulated outbursts with a variety of parameters, for limiting magnitudes set by both the artificial star and background noise techniques for all of our SN~IIn targets.  The results of these tests are shown in Figures \ref{fig:SimIP} and \ref{fig:SimEta}.  We simulate outbursts similar to those from SN 2009ip (many short outbursts) and $\eta$ Carinae (less frequent but extended outbursts) over a range of magnitudes.  Specifically, in the case of  SN 2009ip-like outbursts, we simulate ten-day light curves which vary in peak magnitude between $-13$ and $-16$, and recur between one and ten times in a particular simulation.  In the case of many repeated outbursts, the separation between events is set to ten days.  For instance, a five-event SN 2009ip-like light curve at $m=-15$\,mag uses five separate outburst events that are ten days long at a constant magnitude of $-15$ with ten days separating each of the five outburst events.  In the case of $\eta$ Carinae-like outbursts, we simulate a single light curve which varies in magnitude between $-13$ and $-16$, and duration of the outburst between ten and one hundred days.  Note here that the SN 2008fq and SN 2010jl fractional coverage rates are diminished when the artificial star limiting magnitudes are used because of high host-galaxy brightness in the signal flux aperture.  The remainder of the fractional coverage rates are quite high ($\gtrsim 0.2$) for almost all cases plotted with $M<-13.5\,\mathrm{mag}$.  

The KAIT cadence of 3--10 days is sufficient for the purposes of this project, as can be seen from the varying duration of outburst plots in Figure \ref{fig:SimEta}.  As the duration of the outbursts decreases, the fractional coverage rates drop slowly compared to the steep dropoffs seen in the other plots.  This slow drop is primarily caused by the fact that our limiting magnitudes only cover the $\sim50\%$ of the year during which we have data, so that longer outbursts spend less time in our ``blind spots.''  

Our main point is that the fractional coverage rates become very small for $M_R>-13\,\mathrm{mag}$, indicating that KAIT survey data are not sensitive enough to place constraints on outbursts fainter than this level.  Because the SN 2009ip and $\eta$ Carinae light curves spend most of their quiescent time at $M_R \approx -11\,\mathrm{mag}$ or $M_V \approx -12\,\mathrm{mag}$ (respectively), such sensitivity levels are necessary to thoroughly constrain the frequency of their outbursts.

\subsection{Red vs. Blue Supergiant?}
\label{sec:RvB}
Although it had been speculated for some time that red supergiants (RSGs) were the most likely progenitors of SNe~II, evidence has suggested that some SNe~II may explode while in a BSG or LBV phase.  The most famous of such BSG progenitors is SN 1987A \citep{1989ApJ...343..834A}, but more recent cases associated with SNe IIn have been suggested as well \citep{2010AJ....139.1451S,2011ApJ...732...63S,2014MNRAS.438.1191S}.  The key difference between a RSG and a BSG progenitor is the larger photospheric radius for a RSG progenitor.  Because exploded stars with larger initial radii lose less thermal energy to adiabatic expansion, their SNe can achieve a higher initial peak luminosity \citep[]{1985AJ.....90.2303D}.  Whereas the RSG's light curve essentially plateaus from this peak, a BSG's light curve will continue to rise from its initially faint state as $\mathrm{^{56}Ni}$ decay begins to dominate, as in the example of SN 1987A \citep{1989ApJ...343..834A, 1991ApJ...383..295A}.  

Interaction with CSM can also greatly increase the luminosity from a BSG (or any) explosion depending on the delay between the explosion time and the onset of strong CSM interaction (see \citealp{2014MNRAS.438.1191S}), but in this section we are concerned with the rise to peak brightness of the SN-ejecta photosphere.  Since LBV progenitors to SNe~IIn are expected to have BSG-like radii, we can look for these signatures in the light curve to constrain progenitor properties.  SN 2009ip and SN 2010mc are strong candidates for having relatively compact BSGs prior to explosion (compact, at least, compared to a RSG of a similar luminosity).  Because an abundance of data exists for both the SN explosion itself and the many outbursts in the preceding years, SN 2009ip's light curve is used as a general template to explore our ability to constrain the progenitors for the SNe~IIn studied in this work.  Given that some of our SNe~IIn have limiting magnitudes set shortly before they reached peak brightness, we can use these limits to constrain how quickly the light curve rose in each case.  For brevity, we refer to an initially faint BSG SN light curve as an 09ip/2012a-like event.

\begin{figure*}
\centering
\includegraphics[width=1\textwidth,height=0.95\textheight,clip=true,trim=0cm 0cm 0cm 0cm]{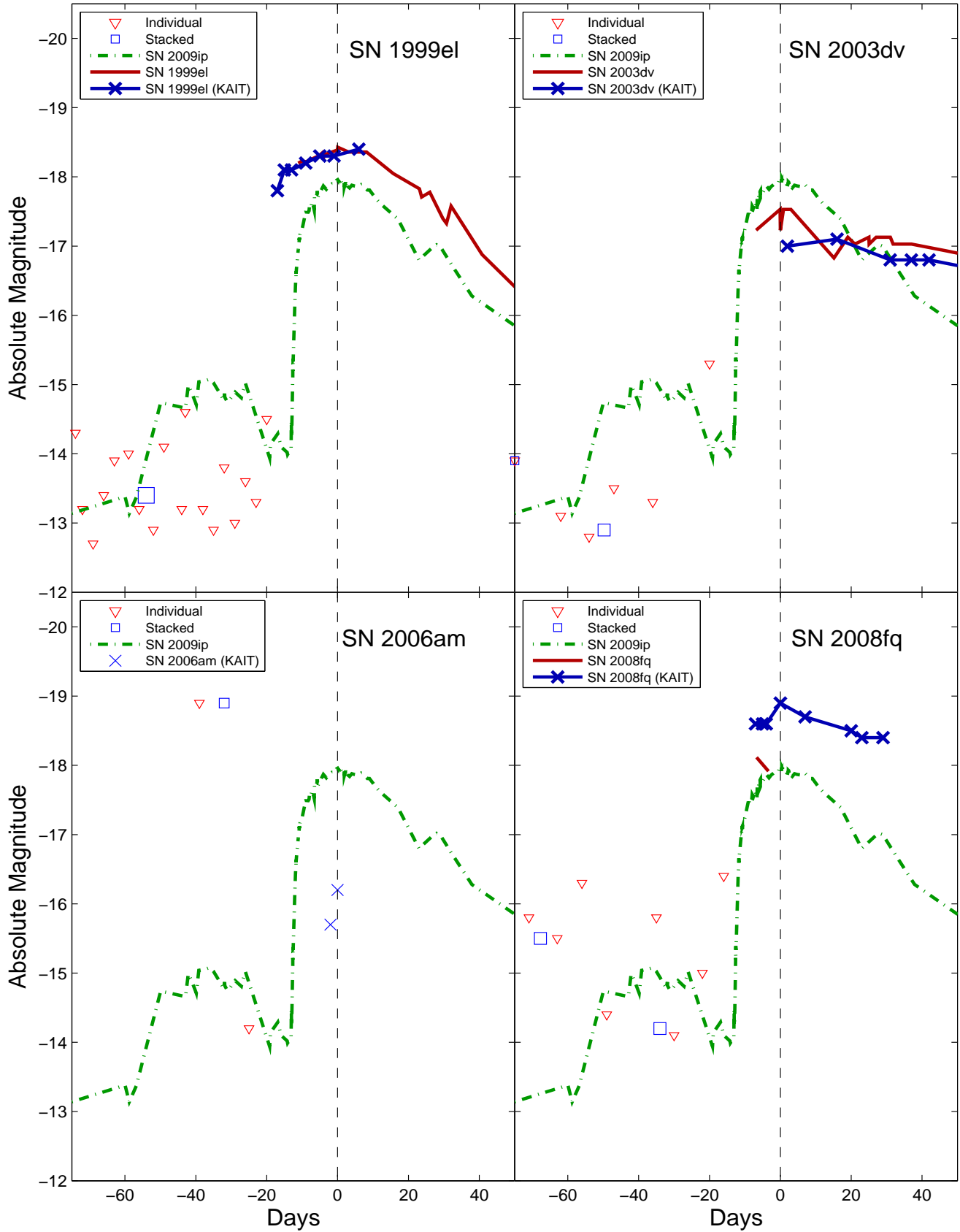}
\caption{Same data format as in Fig. \ref{fig:LM2003dv}, but here we display only zoomed data for the four SNe which have limits within 75 days of their respective peaks.  We compare the data for each SN to those of SN 2009ip and exclude $\eta$ Carinae in this comparison figure.  The data are plotted relative to the observed peak magnitude in published data for each of the SNe if the light curve is well sampled.  If our KAIT light curve provides a better constraint on the date of the peak (as in the case of SN 2008fq), then we use it instead.}
\label{fig:fourfig}
\end{figure*}

SN 1999el, SN 2003dv, SN 2006am, and SN 2008fq have relevant data for this question.  Figure \ref{fig:fourfig} shows zoomed-in light curves for these SNe near their times of peak brightness.  With data ending just 14 days before peak, SN 1999el shows strong signs that it did not have a faint 09ip/2012a-like event prior to reaching its peak brightness.  Data for SN 2003dv are earlier relative to the peak, but do span an effective range of 36 to 75 days prior to the peak brightness, suggesting a possible quick rise to peak as well.  One limit is set 20 days before peak for SN 2003dv, but the noise level is too high to conclusively rule out a 09ip/2012a-like event.  SN 2006am also has limits 25 and 39 days before its observed peak (because we have so few data for the SN 2006am light curve, we do not know for certain that we actually have near-peak observations for this event), but the 25-day limit is from a poor-quality image and does not significantly constrain the SN brightness.  The observation 39 days prior to peak brightness does suggest that no detectable source at $M \approx -14\,\mathrm{mag}$ was present at that time, but this alone does not rule out a 09ip/2012a-like event for SN 2006am.  Pre-SN data near peak for SN 2008fq span a range of 16--135 days prior to the SN explosion, three of which seem to suggest a light curve fainter than that of SN 2009ip as it rose to peak.  However, the majority of the limiting magnitudes are too bright to stringently constrain the rise to peak for SN 2008fq.  This is a direct consequence of its being far away at 47.4 Mpc and being located near the brightest part of its host galaxy.

Because SN 1999el and SN 2003dv do not seem to have had a faint pre-peak SN bump, it is unlikely that their progenitors were BSGs akin to SN 2009ip and SN 2010mc.  Instead, these SNe underwent a quick rise to peak that would be expected from a RSG progenitor, or perhaps from a BSG with immediate onset of strong CSM interaction.  For SN 2010jl and SN 2011A, we cannot place any constraints in this manner because no data are available within 150 days prior to peak brightness, but we do know that {\it HST} images for SN 2010jl suggest a possible BSG progenitor \citep{2011ApJ...732...63S}.  

We must also consider SNe~IIn-P and their progenitors here because a SN~IIn-P may appear as a SN~IIn if no data are available to rule out a plateau phase.  Hence, we cannot exclude the possibility that some of our SNe~IIn are in fact SNe~IIn-P.  This is important because SNe~IIn-P are hypothesised to arise from super asymptotic giant branch (AGB) stars which explode as electron-capture SNe, as in the case of SN 1994W, SN 2009kn, SN 2011ht, and the Crab nebula's SN \citep{2004MNRAS.352.1213C,2012MNRAS.424..855K,2013MNRAS.431.2599M,2013ApJ...779L...8F,2013MNRAS.434..102S}.  SN 2011ht was reported to have an outburst one year prior to its terminal explosion \citep{2013ApJ...779L...8F}, suggesting that super-AGB stars may also produce nonterminal eruptive mass-loss events just before exploding.  The population of events that appear as SNe~IIn may therefore have a diversity of progenitors from RSGs to BSGs rather than a single progenitor system \citep{2009AJ....137.3558S}.

Recent work by \citet{2014arXiv1401.5468O}, which looked at a sample of sixteen SNe~IIn in Palomar Transient Factory (PTF) data, claims that more than 50\% of SNe~IIn have at least one pre-explosion outburst brighter than $3\times10^{7}\,{\rm L}_{\odot}$ (absolute magnitude $M \approx -14$) that occurs within 4 months before the SN.\footnote{This particular statistic assumes a homogeneous population among all of the SN~IIn targets so that it can count multiple precursor events for a single target as essentially different events.  Thus, if sixteen precursor outbursts were detected for a single one of the sixteen different targets, the statistic would claim that essentially 100\% have such an outburst within the measured time frame.}  This statistical estimate includes two events (SN 2010mc precursor 20 days before, and the PTF 12cxj-A precursor 10 days before) that occur just prior to the SN peak brightness, but are very similar to the 09ip/2012a precursor that occurred just 25 days before peak brightness.  \citet{2013MNRAS.434.2721S} showed that the light curves of SN 2010mc and SN 2009ip are almost identical.  These ``precursors'' may well be the initial stages of the slowly rising SN explosion itself before CSM interaction peaks, as was the case for SN 2009ip \citep{2013MNRAS.430.1801M,2014MNRAS.438.1191S}.  For these reasons, it seems that detectable luminous eruptions ($M<-14\,\mathrm{mag}$) may be less common among SN~IIn progenitors just before explosion than the 50\% estimate by \citet{2014arXiv1401.5468O}.  Our constraints on SN 2009ip-like and $\eta$ Carinae-like eruptions suggest that these more luminous eruptive events are uncommon, or that they are fainter than our example outbursts (intrinsically or owing to possible dust extinction).  In fact, we find that our data require that $\lesssim40\%$ of SNe~IIn have pre-SN LBV-like eruptions of $-13$ mag or brighter in the time frame covered by KAIT.

\section{Conclusions}
\label{sec:Con}
Our primary goal was to search through the KAIT archival database in order to constrain the properties of SN~IIn progenitor outbursts by determining if any SNe~IIn exhibit detectable outburst characteristics similar to those of SN 2009ip or $\eta$ Carinae.  No such outbursts were statistically detected in the dataset, implying that they are either typically less luminous than expected or are not common among SNe~IIn.  Instead, we have provided limiting magnitudes for six SN~IIn progenitors.  

With these limiting magnitudes, we are able to place constraints on the nature of a few of the progenitors to our SNe~IIn.  In particular, we find that SN 1999el and SN 2003dv have constraining limiting magnitudes set within 40 days prior to their peak brightness, suggesting that they are likely RSGs with steeply rising light curves or that CSM interaction turned on much more quickly than the delayed interaction in SN 2009ip and SN 2010mc.  We also find that if the SN 2009ip and $\eta$ Carinae outbursts were typical of all SNe~IIn, then we would have a good chance ($>60\%$) of detecting each given type of outburst at least once in our dataset.  However, this is for an emerged luminosity corresponding to $\sim-13\,\mathrm{mag}$.  This estimate ignores the possibility that these objects (as well as SN 2009ip and $\eta$ Carinae) are in fact obscured by their own CSM dust, which is likely to be created if there have been previous such mass-loss episodes.  It also depends somewhat on our particular viewing angle for each SN if the CSM geometry is aspherical \citep{2014MNRAS.442.1166M}.

In order to better study the nature of SNe~IIn and their progenitors, it would be beneficial to perform surveys with deeper limiting magnitudes.  The KAIT cadence of 3--10 days provides excellent temporal coverage of the outbursts we are exploring, while the (up to) 12\,yr baseline of archival data provides nearly unparalleled temporal extent for such a project.  However, the sensitivity of KAIT only allows us to reach limiting magnitudes of $m \approx 19.5$, which restricts us to SNe~IIn within 50\,Mpc.  Any objects at larger distances than this are difficult to constrain because they would likely be too faint for KAIT upper limits to be meaningful.  For example, we searched for pre-SN outbursts in SN 2006gy.  Although we detected no pre-SN outbursts, the upper limit of $M \approx -15\,\mathrm{mag}$ at $d=73.1$\,Mpc is not very constraining.  Ideally, a survey geared toward detecting SN~IIn progenitor outbursts should strive to achieve a limiting magnitude of $m \approx 21.5$ in order to be able to reach sensitivities of $M \approx -12\,\mathrm{mag}$ out to 50\,Mpc.  The Large Synoptic Survey Telescope (LSST) project aims to reach $m \approx 27$\,mag limits in stacked images (mag 24 in single 15\,s exposures).\footnote{Expected limiting magnitude estimates come from \url{http://www.lsst.org/lsst/science/science-faq}.}  At this level of sensitivity, we could measure the progenitors to these SNe~IIn in their relatively quiescent states if they are as luminous as SN 2009ip or $\eta$ Carinae out to $d \approx 500$\,Mpc, but variability is needed to pick them out of the confusion-limited background.  If these objects are not detected by such a survey, then it is likely they are being obscured by circumstellar dust, in which case an infrared survey would be interesting.  Even so, deeper optical surveys would allow us to statistically compare the number of SNe~IIn that arise from objects that undergo tremendous mass loss just prior to explosion.

\section*{Dedication}
This research, based on data from the Lick Observatory Supernova Search that took many years to acquire, was completed after the untimely and tragic death of our close friend, colleague, and coauthor Weidong Li, who led the nightly operation of KAIT (1997--2011) and conducted numerous scientific studies with it; we miss him dearly.

\section*{Acknowledgments}

We thank the Lick Observatory staff for their assistance with the operation of KAIT. We are grateful to the many students, postdocs, and other collaborators who have contributed to KAIT and LOSS over the past two decades, and to discussions concerning the results and SNe in general --- especially S. Bradley Cenko, Ryan Chornock, Ryan J. Foley, Mohan Ganeshalingam, Saurabh W. Jha, Jesse Leaman, Maryam Modjaz, Dovi Poznanski, Frank J. D. Serduke, Jeffrey M. Silverman, Thea Steele, and Xiaofeng Wang.

C.B. is grateful for the Steward Fellowship.  NSF grant AST-1210599 also provided support for this research.  The work of N.S. on $\eta$ Carinae and related outbursts is supported by NSF grant AST-1312221.  The work of A.V.F.'s supernova group at UC Berkeley has been generously supported by the NSF (most recently through grants AST--0908886 and AST-1211916), the TABASGO Foundation, the Christopher R. Redlich Fund, US Department of Energy SciDAC grant DE-FC02-06ER41453, and US Department of Energy grant DE-FG02-08ER41563.  KAIT and its ongoing operation were made possible by donations from Sun Microsystems, Inc., the Hewlett-Packard Company, AutoScope Corporation, Lick Observatory, the NSF, the University of California, the Sylvia \& Jim Katzman Foundation, the Christopher R. Redlich Fund, the Richard and Rhoda Goldman Fund, and the TABASGO Foundation.  We give particular thanks to Russell M. Genet, who made KAIT possible with his initial special gift to A.V.F.; Joseph S. Miller, who allowed KAIT to be placed at Lick Observatory and provided staff support; Jack Borde, who gave invaluable advice regarding the KAIT optics; Richard R. Treffers, KAIT's chief engineer; and the TABASGO Foundation, without which this work would not have been completed. 

This research used astrometric solutions from Astrometry.net and data provided by the USNO-B Image and Catalogue Archive operated by the United States Naval Observatory. We also made use of the NASA/IPAC Extragalactic Database (NED), which is operated by the Jet Propulsion Laboratory, California Institute of Technology, under contract with the National Aeronautics and Space Administration.

\appendix

\bsp

\label{lastpage}

\end{document}